\documentclass[preprint,prd,tightenlines,floatfix,
nofootinbib,eqsecnum,superscriptaddress]{revtex4-1}

\usepackage[T1]{fontenc}		

\usepackage{amsmath,amsfonts,amssymb,amstext,mathrsfs}
\usepackage{mathpazo}

\usepackage[dvips]{graphicx}
\usepackage{epsf,float}
\usepackage{revsymb}

\usepackage{dcolumn}
\usepackage{braket}
\usepackage{color,xcolor}
\usepackage{graphicx}
\usepackage{subfigure}
\usepackage{multirow}
\usepackage{tabularx}
\usepackage{pstricks}
\usepackage[section]{placeins}
\usepackage{booktabs}
\usepackage{array}

\usepackage{tablefootnote}

\usepackage{commath}

\usepackage{hyperref}

\usepackage{lineno}

\newcommand{\Pom}{\mathbb{P}}
\newcommand{\Ode}{\mathbb{O}}
\newcommand{\Reg}{\mathbb{R}}

\newcommand{\bptap}{\mbox{\boldmath $p'_{t,1}$}}
\newcommand{\bptbp}{\mbox{\boldmath $p'_{t,2}$}}

\newcommand{\bk}{\mbox{\boldmath $k$}}
\newcommand{\bpa}{\mbox{\boldmath $p_{a}$}}

\newcommand{\bpaap}{\mbox{\boldmath $p_{1}'$}}

\renewcommand\slash[1]{\not \! #1}

\usepackage[normalem]{ulem}  

\begin{document}

\title{\boldmath 
Central exclusive diffractive production of a single photon
in high-energy proton-proton collisions
within the tensor-Pomeron approach}

\vspace{0.6cm}

\author{Piotr Lebiedowicz}
\email{Piotr.Lebiedowicz@ifj.edu.pl}
\affiliation{Institute of Nuclear Physics Polish Academy of Sciences, Radzikowskiego 152, PL-31342 Krak{\'o}w, Poland}

\author{Otto Nachtmann}
\email{O.Nachtmann@thphys.uni-heidelberg.de}
\affiliation{Institut f\"ur Theoretische Physik, Universit\"at Heidelberg,
Philosophenweg 16, D-69120 Heidelberg, Germany}

\author{Antoni Szczurek
\footnote{Also at \textit{College of Natural Sciences, 
Institute of Physics, University of Rzesz{\'o}w, 
ul. Pigonia 1, PL-35310 Rzesz{\'o}w, Poland}.}}
\email{Antoni.Szczurek@ifj.edu.pl}
\affiliation{Institute of Nuclear Physics Polish Academy of Sciences, Radzikowskiego 152, PL-31342 Krak{\'o}w, Poland}

\begin{abstract}
We discuss central-exclusive production (CEP) of photons
via different fusion processes in the reaction $pp \to pp \gamma$
at high energies, available at RHIC and LHC,
within the tensor-pomeron model.
We consider two types of processes, the photoproduction contribution
via the photon-pomeron and photon-reggeon fusion reactions,
and the purely diffractive contribution
via the reggeon-pomeron and odderon-pomeron fusion reactions.
We present predictions for the measurements of photons 
at midrapidity, $|{\rm y}| < 2.5$,
and at relatively low transverse momentum,
$0.1~{\rm GeV} < k_{\perp} < 1~{\rm GeV}$.
To check the main results of our study 
the measurement of the outgoing protons is not necessary.
This is of relevance, e.g., for the present version
of the ALICE detector at the LHC.
Several differential distributions,
for instance, in ${\rm y}$, $k_{\perp}$ and $\omega$,
the rapidity, the absolute value of the transverse momentum,
and the energy of the photon, respectively, are presented.
We show that the photoproduction 
is an important process in the kinematic region specified above.
There it gives a much larger cross section
than diffractive bremsstrahlung where 
the basic $pp \to pp$ reaction is due to strong interaction diffraction.
This is remarkable as the CEP cross section is of order
$\alpha_{\rm em}^{3}$ whereas the bremsstrahlung one is
only of order $\alpha_{\rm em}$.
On the other hand, the soft-photon bremsstrahlung
is more important than CEP
in the forward rapidity range, $|{\rm y}| > 4$,
and/or at very low $k_{\perp}$.
We leave it as a challenge for the planned ALICE 3 experiment 
at the LHC to study these two contributions to soft photon production
in $pp$ collisions. This could shed new light on the so-called
``soft photon puzzle'' in hadron-hadron collisions.
\end{abstract}


\maketitle

\section{Introduction}
\label{sec:1}

In this article we continue our investigations
of exclusive photon production
in high-energy hadronic collisions
in the tensor-pomeron approach.
In \cite{Lebiedowicz:2021byo} 
we have treated $\pi \pi$ scattering 
without and with photon radiation.
In \cite{Lebiedowicz:2022nnn} we have discussed
the soft-photon bremsstrahlung 
in the $pp \to pp \gamma$ reaction.
In the present paper we extend our considerations
to central-exclusive production (CEP) processes 
of single photon in high-energy proton-proton collisions.

The $pp \to pp \gamma$ reaction
was not yet measured at high energies.
There is, however,
a plan for a new multipurpose detector at the LHC, 
ALICE~3 \cite{Adamova:2019vkf,QM2022_PBM,EMMI_RRTF,ALICE:2022wwr},
that would be able to measure
ultra-soft photons at very low transverse momentum in 
$pp$, $pA$, and $AA$ collisions.
The main aim of our paper is to discuss 
CEP processes in the exclusive $pp \to pp \gamma$ reaction 
at low transverse momentum of the photon for the LHC energy range.

A measurement of the soft-photon production at the LHC
could shed light on a long-standing discrepancy between
the theoretical predictions of the bremsstrahlung,
based on Low's theorem \cite{Low:1958sn}, 
and the measured soft-photon spectra in several hadronic reactions.
For experiments on soft photon production see \cite{Goshaw:1979kq,Chliapnikov:1984ed,Botterweck:1991wf,Banerjee:1992ut,Antos:1993wv,Tincknell:1996ks,Belogianni:1997rh,Belogianni:2002ib,Belogianni:2002ic,Abdallah:2005wn,Abdallah:2007aa,DELPHI:2010cit}.
Overviews of the experimental and theoretical status of this
``soft photon puzzle'' are given in \cite{Wong:2014pY} and \cite{QM2022_PBM}.
The question is if there is production of so called
``anomalous soft photons'', and if so, what is the origin of these photons.
From our point of view the origin of such anomalous terms
should be searched for in nonperturbative QCD processes.
In this paper we will consider 
a conventional source of ``anomalous photons'', that is,
photons from CEP reactions.
For unconventional sources of anomalous photons
see e.g. \cite{Botz:1994bg,Nachtmann:2014qta} 
and the review in \cite{Wong:2014pY}.

In order to calculate the cross section for $pp \to pp \gamma$
we use the tensor-pomeron and vector-odderon model 
proposed in \cite{Ewerz:2013kda}.
In this model, the charge-conjugation $C = +1$ exchanges, that is,
the pomeron $\Pom$ and the reggeons
$\Reg_{+} = f_{2 \Reg}$, $a_{2 \Reg}$
are treated as effective rank-2 symmetric tensor exchanges,
the $C = -1$ odderon $\Ode$ and the reggeons 
$\Reg_{-} = \omega_{\Reg}$, $\rho_{\Reg}$ 
are described as effective vector exchanges.
In \cite{Ewerz:2016onn} the helicity structure of
high-energy $pp$ elastic scattering 
at small momentum transfers was calculated.
It was shown there that the STAR data \cite{Adamczyk:2012kn} 
exclude a scalar character of the pomeron-proton coupling but
are perfectly compatible with the tensor-pomeron model.
The assumption of a vector character for the pomeron couplings
has theoretical and experimental problems as discussed in \cite{Ewerz:2016onn,Britzger:2019lvc}.
In \cite{Britzger:2019lvc} it was shown that 
for a pomeron coupling to photons like a vector
its contribution 
to real Compton scattering
and hence to the total $\gamma p$ photoabsorption 
cross section vanishes exactly.
A further result of \cite{Britzger:2019lvc} is that
a vector pomeron cannot contribute to the forward virtual Compton amplitude which leads to the conclusion
of its decoupling in the structure functions of low-$x$
deep inelastic scattering (DIS).
On the other hand, the data for the total $\gamma p$ photoabsorption
cross section and the \mbox{low-$x$} structure functions
clearly indicate that at high energies pomeron exchange
must be present as a \textit{main} contribution.
And indeed, in the tensor-pomeron model a very satisfactory fit
of these data was obtained; see \cite{Britzger:2019lvc}.
Applications of the tensor-pomeron concept
were given for photoproduction of pion pairs in \cite{Bolz:2014mya}
and for a number of diffractive CEP reactions 
in $pp$ collisions at high energies
\cite{Lebiedowicz:2013ika,Lebiedowicz:2014bea,
Lebiedowicz:2016ioh,Lebiedowicz:2016ryp,Lebiedowicz:2016zka,
Lebiedowicz:2018sdt,Lebiedowicz:2018eui,
Lebiedowicz:2019boz,Lebiedowicz:2019jru,
Lebiedowicz:2019por,Lebiedowicz:2020yre,
Lebiedowicz:2021pzd}.

Several processes contribute to the $pp \to pp \gamma$ reaction.
One of them is CEP of single photons through 
the $\gamma - \Pom$-fusion process.
In order to calculate the relevant amplitudes 
we need the $\Pom \gamma \gamma$ coupling functions.
In addition to the $\gamma - \Pom$-fusion process
we shall also estimate the subleading 
$\gamma - f_{2 \Reg}$-fusion process.
The \textit{Ans\"atze} for the relevant vertices
for real and virtual photons
are discussed in \cite{Ewerz:2013kda,Britzger:2019lvc,Lebiedowicz:2022xgi}.
The $Q^{2}$ and $t$ dependencies of the coupling functions
in the pomeron/reggeon-photon-photon vertices must be
determined from a comparison to experimental data.
We shall use the parametrizations obtained in \cite{Lebiedowicz:2022xgi}
from a comparison of the tensor-pomeron model 
to the elastic $\gamma p$-scattering data from FNAL and
to the deeply virtual Compton scattering (DVCS) data 
measured at HERA.
The CEP of photons via the $\gamma - \Pom$-fusion process
can be expected to populate preferentially 
the midrapidity region as was discussed 
earlier in \cite{Lebiedowicz:2013xlb} within another approach.
In the present paper, we shall also discuss CEP of single photons
through the $\Reg_{-} - \Pom$, $\Reg_{-} - \Reg_{+}$,
$\Ode - \Pom$, and $\Ode - \Reg_{+}$ fusion processes 
within the tensor-pomeron and vector-odderon approach \cite{Ewerz:2013kda}.
We wish to estimate the size of the cross sections 
for these processes for the LHC energy range.
In a first approximation we neglect absorption effects 
due to the proton-proton interactions.

Our paper is organized as follows.
In the next section we give analytic expressions 
for the amplitudes for the $pp \to pp \gamma$ reaction.
Different CEP fusion processes such as
$\gamma - \Pom$, $\rho_{\Reg} - \Pom$, and $\Ode - \Pom$
are discussed.
The results of our calculations are presented in Sec.~\ref{sec:3}.
Section~\ref{sec:4} contains a summary and our conclusions.
In Appendix~\ref{sec:appendixA} we list
the expressions for the effective propagators and vertices
used in our model.
Appendix~\ref{sec:appendixB} is devoted to an approximate
calculation of $pp \to pp \gamma$ using
the method of the equivalent photon spectrum.
This gives us an understanding of the size of the cross sections
obtained with our model.
In Appendix~\ref{sec:appendixC} 
we discuss the CEP mechanism for photons
in the soft photon limit $k \to 0$, 
where $k$ is the photon's four-momentum.

\section{Theoretical formalism}
\label{sec:2}

We consider the reaction
\begin{eqnarray}
p (p_{a},\lambda_{a}) + p (p_{b},\lambda_{b}) \to 
p (p_{1}',\lambda_{1}) + p (p_{2}',\lambda_{2}) + \gamma(k, \epsilon)\,.
\label{pp_ppgam}
\end{eqnarray}
The momenta are indicated in brackets,
the helicities of the protons are denoted by
$\lambda_{a}, \lambda_{b}, \lambda_{1}, \lambda_{2} 
\in \{1/2, -1/2 \}$,
and $\epsilon$ is the polarization vector of the photon.

The kinematic variables are
\begin{eqnarray}
&&s = (p_{a} + p_{b})^{2} = (p_{1}' + p_{2}' + k)^{2}\,, \nonumber \\
&& q_{1} = p_{a} - p_{1}' \,, 
\quad t_{1} = q_{1}^{2}\,, \nonumber \\
&& q_{2} = p_{b} - p_{2}' \,,
\quad t_{2} = q_{2}^{2}\,, \nonumber \\
&& s_{1} = W_{1}^{2} = (p_{1}' + k)^{2} = (p_{a} + q_{2})^{2}\,,
\nonumber \\
&& s_{2} = W_{2}^{2} = (p_{2}' + k)^{2} = (p_{b} + q_{1})^{2}\,,
\nonumber \\
&& u_{1} = (p_{a} - p_{2}')^{2}\,,
\nonumber \\
&& u_{2} = (p_{b} - p_{1}')^{2}\,.
\label{2.17}
\end{eqnarray}
In the overall c.m. system we choose the 3 axis in the direction of $\bpa$.
The rapidity of the photon is then
\begin{eqnarray}
{\rm y} = \frac{1}{2} \ln \frac{k^{0} + k^{3}}{k^{0} - k^{3}} = - \ln \tan \frac{\theta}{2}\,,
\label{y}
\end{eqnarray}
where $\theta$ is the polar angle of $\bk$, $k^{3} = |\bk| \cos\theta$.
Furthermore, 
$\omega = k^{0}$ is
the energy of the photon,
$\omega = |\bk_{\perp}| \cosh \rm{y}$.

The ${\cal T}$-matrix element for the reaction (\ref{pp_ppgam}) is
\begin{eqnarray}
&&\braket{p(p_{1}',\lambda_{1}),p(p_{2}',\lambda_{2}),
\gamma(k, \epsilon)|{\cal T}|p(p_{a},\lambda_{a}),p(p_{b},\lambda_{b})}  \nonumber\\
&&\qquad = 
(\epsilon^{\mu})^{*}
{\cal M}^{(\rm total)}_{\mu}(p_{a},\lambda_{a}; p_{b},\lambda_{b};
p_{1}',\lambda_{1};p_{2}',\lambda_{2};k)\,.
\label{2.18}
\end{eqnarray}
The amplitude must be antisymmetric
under interchange of the two final protons
\begin{equation}
{\cal M}^{(\rm total)}_{\mu}(p_{a},\lambda_{a}; p_{b},\lambda_{b};
p_{1}',\lambda_{1};p_{2}',\lambda_{2};k) = 
-{\cal M}^{(\rm total)}_{\mu}(p_{a},\lambda_{a}; p_{b},\lambda_{b};
p_{2}',\lambda_{2};p_{1}',\lambda_{1};k)
\label{2.20a}
\end{equation}
and gauge invariance requires
\begin{equation}
k^{\mu} {\cal M}^{(\rm total)}_{\mu} = 0\,.
\label{2.19}
\end{equation}
We are interested in high c.m. energies $\sqrt{s}$ 
and small momentum transfers $|t_{1}|$, $|t_{2}|$:
\begin{equation}
\sqrt{s} \gg m_{p}\,, \quad 
|t_{1}|, |t_{2}| \lesssim c m_{p}^{2}\,, \quad 
c = {\cal O}(1)\,.
\label{2.5.1}
\end{equation}
In this kinematic region the amplitude (\ref{2.18}) is
governed by $t$-channel exchanges.
Let us denote the corresponding amplitude by
\begin{equation}
{\cal M}^{(t-{\rm channel})}_{\mu}(p_{a},\lambda_{a}; p_{b},\lambda_{b};
p_{1}',\lambda_{1};p_{2}',\lambda_{2};k)\,.
\label{2.5.2}
\end{equation}
With the exchange of the final-state protons we get
the $u$-channel exchange amplitude,
\begin{equation}
{\cal M}^{(u-{\rm channel})}_{\mu} =
{\cal M}^{(t-{\rm channel})}_{\mu}\mid_{(p_{1}',\lambda_{1})\leftrightarrow(p_{2}',\lambda_{2})}\,.
\label{2.5.3}
\end{equation}
The total amplitude for (\ref{pp_ppgam}) is then
\begin{equation}
{\cal M}^{(\rm total)}_{\mu} =
{\cal M}^{(t-{\rm channel})}_{\mu} -
{\cal M}^{(u-{\rm channel})}_{\mu}\,.
\label{2.5.4}
\end{equation}
In the kinematic region (\ref{2.5.1}) the $u$-channel-exchange term
on the r.h.s. of (\ref{2.5.4}) is expected to give negligible contribution.
Therefore, in the following we omit 
the term ${\cal M}^{(u-{\rm channel})}_{\mu}$
in our considerations
and, for brevity of notation, we set 
\begin{equation}
{\cal M}_{\mu} \equiv {\cal M}^{(t-{\rm channel})}_{\mu}\,.
\label{2.9a}
\end{equation}

As discussed in Sec.~II~C of \cite{Lebiedowicz:2022nnn}
the cross section for the photon yield can then be calculated as follows
\begin{eqnarray}
d\sigma({pp \to pp \gamma}) &=&
\frac{1}{2\sqrt{s(s-4 m_{p}^{2})}}
\frac{d^{3}k}{(2 \pi)^{3} \,2 k^{0}}
\int 
\frac{d^{3}p_{1}'}{(2 \pi)^{3} \,2 p_{1}'^{0}}
\frac{d^{3}p_{2}'}{(2 \pi)^{3} \,2 p_{2}'^{0}}
\nonumber \\
&&\times 
(2 \pi)^{4} \delta^{(4)}(p_{1}'+p_{2}'+k-p_{a}-p_{b})
\frac{1}{4}
\sum_{p\; {\rm spins}}
{\cal M}_{\mu} {\cal M}_{\nu}^{*} (-g^{\mu \nu})\,; \quad 
\label{xs_2to3}
\end{eqnarray}
see Eq.~(2.35) of \cite{Lebiedowicz:2022nnn}.

In the calculations we can consider the amplitude ${\cal M}_{\mu}$ (\ref{2.9a})
as a sum due to the bremsstrahlung (BS)
and the CEP processes contributing to $pp \to pp \gamma$:~\footnote{In
the general case a strict separation
of bremsstrahlungs and CEP contributions is not possible;
see the discussion in Appendix~\ref{sec:appendixC}.}
\begin{equation}
{\cal M}_{\mu} = {\cal M}_{\mu}^{(\rm BS)}
+ {\cal M}_{\mu}^{(\rm CEP)}\,.
\label{amp_total}
\end{equation}
The amplitude ${\cal M}_{\mu}^{(\rm BS)}$
corresponds to diffractive bremsstrahlung
discussed in~\cite{Lebiedowicz:2022nnn};
see the diagrams (a)--(f) of Fig.~3 there.
In this mechanism, the amplitudes corresponding 
to photon emission from
the external protons are determined by the off-shell
$pp$ elastic scattering amplitude and the contact terms
needed in order to satisfy gauge-invariance constraints.
For details how to calculate the bremsstrahlung contribution
in our approach
we refer the reader to Sec.~II and 
Appendix~B of \cite{Lebiedowicz:2022nnn}.

The amplitude for central-exclusive production (CEP) of photons
is given by the sum of the contributions from
the relevant fusion processes
\begin{equation}
{\cal M}_{\mu}^{(\rm CEP)}
= 
{\cal M}_{\mu}^{(\gamma-\Pom)} + {\cal M}_{\mu}^{(\gamma-\Reg_{+})} +
{\cal M}_{\mu}^{(\Reg_{-}-\Pom)} + {\cal M}_{\mu}^{(\Reg_{-}-\Reg_{+})} +
{\cal M}_{\mu}^{(\Ode-\Pom)} + {\cal M}_{\mu}^{(\Ode-\Reg_{+})} \,.
\label{amp_CEP}
\end{equation}
Here, $\Reg_{+}$ denotes the $C = +1$ reggeons 
($f_{2 \Reg}, a_{2 \Reg}$),
and $\Reg_{-}$ denotes the $C = -1$ reggeons
($\omega_{\Reg},\rho_{\Reg}$).
Different fusion processes should be considered,
these involving the photon, 
$\gamma - \Pom$ and $\gamma - \Reg_{+}$, 
as well as purely diffractive contributions
$\Reg_{-} - \Pom$, $\Reg_{-} - \Reg_{+}$, $\Ode - \Pom$, and $\Ode - \Reg_{+}$.
In fact, we consider the exchange of soft ($\Pom_{1}$) 
and hard ($\Pom_{0}$) pomeron. 
Thus, $\Pom$ in (\ref{amp_CEP})
stands for the sum of $\Pom_{1}$ and $\Pom_{0}$,
$\Reg_{+}$ for $f_{2 \Reg}$ and $a_{2 \Reg}$,
and $\Reg_{-}$ for $\omega_{\Reg}$ and $\rho_{\Reg}$.

It is interesting to list the leading order in 
$e = \sqrt{4 \pi \alpha_{\rm em}}$
with which the various processes contribute to ${\cal M}_{\mu}$;
see Table~\ref{Table1}.
Thus, naively one could expect that the fusion processes
$\gamma - \Pom$ and $\gamma - \Reg_{+}$
give small contributions since they are of higher order in $e$,
compared e.g. to diffractive bremsstrahlung.
But, as we shall see, in certain regions of phase space these
$e^{3}$ processes are the dominant ones.
\begin{table}
\caption{The leading order in $e$ of the various processes contributing
to $pp \to pp \gamma$; see (\ref{amp_total}) and (\ref{amp_CEP}).
By diffractive and QED bremsstrahlung we denote the processes
where the basic $pp \to pp$ reaction is due to strong interaction
diffraction and exchange of a photon, respectively.}
\begin{center}
\begin{tabular}{|l|c|}
\hline
Process                             & Leading order in $e$ \\ \hline 
Diffractive bremsstrahlung          & $e$                  \\ 
QED bremsstrahlung                  & $e^{3}$                  \\ 
Fusion $\gamma - \Pom$, $\gamma - \Reg_{+}$  & $e^{3}$              \\ 
Fusion $\Ode - \Pom$, $\Ode - \Reg_{+}$, 
$\Reg_{-} - \Pom$, $\Reg_{-} - \Reg_{+}$     & $e$                  \\ \hline
\end{tabular}
\end{center}
\label{Table1}
\end{table}

In the following, we will discuss the CEP contributions (\ref{amp_CEP})
in detail.

\subsection{Photoproduction contributions}
\label{sec:2B}

\begin{figure}[!h]
(a)\includegraphics[width=6cm]{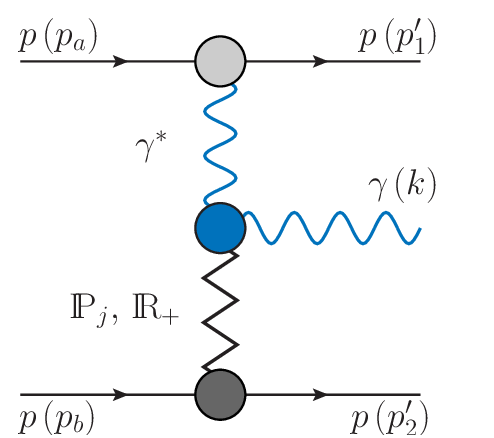}\qquad
(b)\includegraphics[width=6cm]{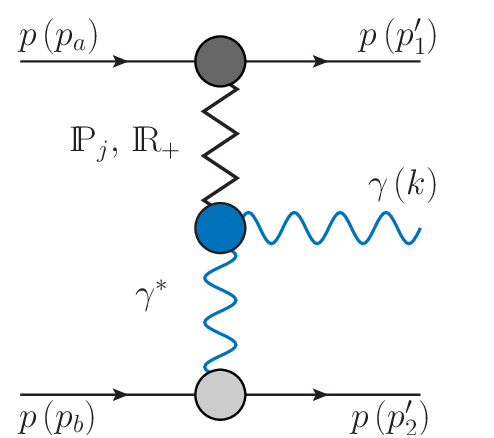}
\caption{Diagrams for CEP of a photon in high-energy proton-proton collisions:
(a) photon-pomeron/reggeon fusion;
(b) pomeron/reggeon-photon fusion.
We have $j = 0,1$ with $\Pom_{0}$ and $\Pom_{1}$ denoting
the hard and soft pomeron, respectively, and $\Reg_{+}$ stands 
for the sum of $f_{2 \Reg}$ and $a_{2 \Reg}$
reggeons.}
\label{fig:pp_pp_gam_CEP_photoprod}
\end{figure}

First we consider the fusion processes involving photon
exchange, $\gamma - \Pom$ and $\gamma - \Reg_{+}$.
The corresponding diagrams
are shown in Fig.~\ref{fig:pp_pp_gam_CEP_photoprod}.
We have for the $\gamma - \Pom$ contribution in Eq.~(\ref{amp_CEP})
\begin{eqnarray}
{\cal M}_{\mu}^{(\gamma - \Pom)} = {\cal M}_{\mu}^{(\gamma \Pom)} + 
{\cal M}_{\mu}^{(\Pom \gamma)}\,.
\label{gamP}
\end{eqnarray}
Here and in the following we indicate by $(\gamma \Pom)$
the amplitude obtained from to the diagram
Fig.~\ref{fig:pp_pp_gam_CEP_photoprod}(a)
and by $(\Pom \gamma)$ that obtained from
Fig.~\ref{fig:pp_pp_gam_CEP_photoprod}(b).
The sum of these two amplitudes is indicated by $(\gamma-\Pom)$.
For other CEP processes we use the analogous notation.

The lower part of the diagram 
in Fig.~\ref{fig:pp_pp_gam_CEP_photoprod}(a)
corresponds exactly to the DVCS diagram 
for $\gamma^{*} p \to \gamma p$ at a reasonably 
high c.m. energy $W_{2}$
such that the $t$-channel exchanges dominate.
In \cite{Lebiedowicz:2022xgi} we have studied this reaction
in the tensor-pomeron-approach and we obtained a good description
of the data for large c.m. energies and small Bjorken $x$ values.
This is the kinematic region where our model should be valid.
How can we assure for the reaction (\ref{pp_ppgam}),
which we study in the present paper,
that we are in a kinematic region where our model is valid?
Let us consider first the diagram of 
Fig.~\ref{fig:pp_pp_gam_CEP_photoprod}(a).
In the lower part of the diagram, the DVCS part,
we should thus have high enough energy, $W_{2} \gtrsim 6$~GeV say,
and small Bjorken $x$ corresponding to 
$|t_{1}|/W_{2}^{2} \ll 1$.
For the diagram of Fig.~\ref{fig:pp_pp_gam_CEP_photoprod}(b)
the analogous conditions are $W_{1} \gtrsim 6$~GeV
and $|t_{2}|/W_{1}^{2} \ll 1$.
However, imposing such conditions by hand would be inconvenient
from an experimental point of view,
since it would require the measurement of the final state protons.
Instead we shall only impose conditions on the photon kinematics.
We shall require the photon transverse momentum $k_{\perp}$
to be in the range $0.1~{\rm GeV} < k_{\perp} < 1$~GeV
and we shall require large rapidity gaps between the centrally
produced photon and the outgoing protons, that is, we
require for the photon rapidity $|{\rm y}| < 2.5$.
The large rapidity gaps will assure dominance of pomeron exchange
in the diagrams of Fig.~\ref{fig:pp_pp_gam_CEP_photoprod}.
We shall see below in Sec.~\ref{sec:3} that, indeed, these requirements
assure that we can safely use 
the diagrams of Fig.~\ref{fig:pp_pp_gam_CEP_photoprod}
for the calculation of the amplitudes for (\ref{pp_ppgam}).

Some comments are in order here.
If we consider instead of CEP of a photon the CEP of a meson $M$
of mass $m_{M}$ the situation is quite different.
Let us, for instance, consider the production of the meson $M$
at rest in the overall c.m. system. We have then automatically
large subenergies squared $W_{1}^{2}, W_{2}^{2} \approx m_{M} \sqrt{s}$
and the exchange diagrams analogous 
to Figs.~\ref{fig:pp_pp_gam_CEP_photoprod}(a) and
\ref{fig:pp_pp_gam_CEP_photoprod}(b) should be valid
representations of the amplitude.
This argument clearly fails for the massless photon.
But, as we shall see in Sec.~\ref{sec:3}, considering
CEP of the photon at midrapidity region
and at low transverse momentum already
ensures large enough energies $W_{1,2}$.
Finally we note that we could extend 
our calculation of CEP of $\gamma(k)$
with $\gamma^{*}$ exchange to a larger region of phase space
using representations of the DVCS data valid 
for higher values of Bjorken $x$ and smaller energies $W_{1,2}$.
For some remarks on this problem see Appendix~\ref{sec:appendixC}.
But a complete discussion of such an extended calculation
goes beyond the scope of our present paper.

Now we come back to the calculation of the diagrams of Fig.~\ref{fig:pp_pp_gam_CEP_photoprod}. 
We consider there the exchanges of $\Pom_{1}, \Pom_{0}$, and $\Reg_{+}$
that correspond to the soft pomeron, the hard pomeron, 
and the reggeons ($f_{2 \Reg} + a_{2 \Reg}$), respectively.
The \textit{Ans\"atze} for effective propagator 
and vertex functions of these exchanges
are taken from \cite{Ewerz:2013kda,Britzger:2019lvc}
and are discussed in Appendix~\ref{sec:appendixA}.
In the tensor pomeron model the pomeron- and reggeon-$\gamma \gamma$ vertices 
have two coupling functions; see (\ref{A3}).
By comparing the tensor-pomeron model
and the experimental data on real Compton scattering from FNAL and on DVCS
obtained by the H1 and ZEUS Collaborations at HERA
we fixed in \cite{Lebiedowicz:2022xgi} these coupling functions.
In our calculations we shall use the FIT~2 parametrization
from \cite{Lebiedowicz:2022xgi} 
[see (\ref{FIT2}) in Appendix~\ref{sec:appendixA}].

The $\gamma \Pom$-exchange amplitude  
can now be written as
\begin{eqnarray}
{\cal M}_{\mu}^{(\gamma \Pom)}&=& (-i) \sum_{j = 0,1}
\bar{u}(p_{1}',\lambda_{1}) i \Gamma^{(\gamma pp)}_{\nu_{1}}(p_{1}',p_{a}) u(p_{a},\lambda_{a})\,
i \Delta^{(\gamma) \,\nu_{1} \nu}(q_{1})\,
i \Gamma^{(\Pom_{j} \gamma^{*} \gamma^{*})}_{\mu \nu \kappa \rho}(k,q_{1})\nonumber\\
&& \times 
i \Delta^{(\Pom_{j}) \,\kappa \rho,\alpha \beta}(s_{2},t_{2}) 
\bar{u}(p_{2}',\lambda_{2}) i \Gamma^{(\Pom_{j} pp)}_{\alpha\beta}(p_{2}',p_{b}) u(p_{b},\lambda_{b})
\nonumber\\
&=&
i \sum_{j = 0,1}
\bar{u}(p_{1}',\lambda_{1}) \Gamma^{(\gamma pp)\,\nu}(p_{1}',p_{a}) u(p_{a},\lambda_{a})\,
 \frac{1}{t_{1}}\,
\frac{1}{2 s_{2}}
\big(-i s_{2} \alpha_{\Pom_{j}}'\big)^{\alpha_{\Pom_{j}}(t_{2})-1}
\nonumber\\
&&\times 
\left[2 a_{\Pom_{j} \gamma^{*} \gamma^{*}}(t_{1},k^{2},t_{2})
\Gamma_{\mu \nu}^{(0)\, \alpha \beta}(k,-q_{1})
-b_{\Pom_{j} \gamma^{*} \gamma^{*}}(t_{1},k^{2},t_{2})
\Gamma_{\mu \nu}^{(2)\, \alpha \beta}(k,-q_{1}) \right]
\nonumber\\
&&\times 
\bar{u}(p_{2}',\lambda_{2}) \Gamma^{(\Pom_{j} pp)}_{\alpha \beta}(p_{2}',p_{b}) u(p_{b},\lambda_{b}).
\label{gamP_amp}
\end{eqnarray}
We use the standard $\gamma$ propagator and the $\gamma pp$ vertex,
see (3.1) and (3.26)--(3.32) of \cite{Ewerz:2013kda}, respectively.
All effective propagator and vertex functions 
for the $\Pom_{j}$ exchanges used in (\ref{gamP_amp})
are listed in Appendix~\ref{sec:appendixA}.

For the $\Pom \gamma$-exchange amplitude
[see Fig.~\ref{fig:pp_pp_gam_CEP_photoprod}(b)]
we have the same structure 
as for the amplitude (\ref{gamP_amp}) with the replacements
\begin{eqnarray}
(p(p_{a}, \lambda_{a}),p(p_{1}',\lambda_{1}))
\leftrightarrow
(p(p_{b}, \lambda_{b}),p(p_{2}',\lambda_{2}))\,,\quad 
t_{1} \leftrightarrow t_{2}\,,\quad 
q_{1} \leftrightarrow q_{2}\,,\quad
s_{2} \leftrightarrow s_{1}\,.\qquad
\label{replacement}
\end{eqnarray}
In a similar way we obtain the $\gamma \Reg_{+}$- 
and $\Reg_{+} \gamma$-exchange amplitudes.

\subsection{Diffractive contributions}
\label{sec:2C}

\begin{figure}[!h]
(a)\includegraphics[width=6cm]{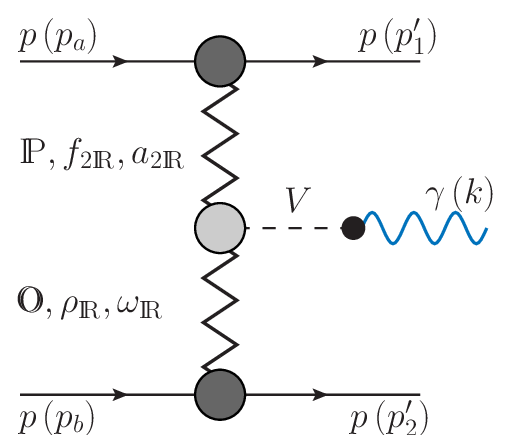}\qquad
(b)\includegraphics[width=6cm]{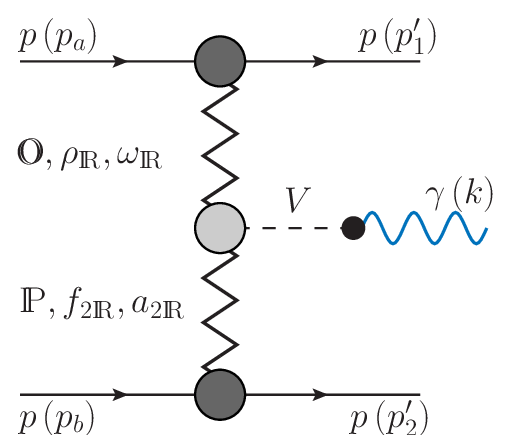}
\caption{
Diagrams for diffractive production of a photon
in high-energy proton-proton collisions
via reggeon, pomeron, and odderon exchanges:
(a) ($\Reg_{+}$/pomeron)-($\Reg_{-}$/odderon) fusion;
(b) ($\Reg_{-}$/odderon)-($\Reg_{+}$/pomeron) fusion.
We use here the vector-meson dominance (VMD) approach
and $V$ stands for the appropriate vector meson
$\rho^{0}, \omega, \phi$ which is coupling 
to the photon according to (\ref{2.15.1}).}
\label{fig:pp_pp_gam_CEP}
\end{figure}

Here we consider the contributions
from purely diffractive fusion processes
given by the diagrams of Fig.~\ref{fig:pp_pp_gam_CEP}.
We have the following fusion processes (see (\ref{amp_CEP})):
\begin{eqnarray}
&&{\cal M}_{\mu}^{(\Reg_{-}-\Pom)} = 
{\cal M}_{\mu}^{(\Pom \rho_{\Reg})} + 
{\cal M}_{\mu}^{(\rho_{\Reg} \Pom)} +
{\cal M}_{\mu}^{(\Pom \omega_{\Reg})} + 
{\cal M}_{\mu}^{(\omega_{\Reg} \Pom)} \,,
\label{PR}\\
&&{\cal M}_{\mu}^{(\Reg_{-}-\Reg_{+})} =
{\cal M}_{\mu}^{(f_{2 \Reg} \rho_{\Reg})} + 
{\cal M}_{\mu}^{(\rho_{\Reg} f_{2 \Reg})} +
{\cal M}_{\mu}^{(f_{2 \Reg} \omega_{\Reg})} + 
{\cal M}_{\mu}^{(\omega_{\Reg} f_{2 \Reg})} +
(f_{2 \Reg} \to a_{2 \Reg})\,,
\label{RR}\\
&&{\cal M}_{\mu}^{(\Ode - \Pom)} = 
  {\cal M}_{\mu}^{(\Pom \Ode)} + 
  {\cal M}_{\mu}^{(\Ode \Pom)}\,,
\label{PO}\\
&&{\cal M}_{\mu}^{(\Ode - \Reg_{+})} = 
  {\cal M}_{\mu}^{(f_{2 \Reg} \Ode)} + 
  {\cal M}_{\mu}^{(\Ode f_{2 \Reg})} +
  (f_{2 \Reg} \to a_{2 \Reg})\,.
\label{OR}
\end{eqnarray}
For these diffractive fusion processes we assume that only the soft pomeron $\Pom_{1}$ contributes.
At high c.m. energies $\sqrt{s}$ 
the $\Reg_{-}-\Reg_{+}$-fusion processes (\ref{RR})
can be safely neglected.

As indicated in Fig.~\ref{fig:pp_pp_gam_CEP} we use here the vector-meson
dominance (VMD) approach.
We assume that an appropriate vector meson $V$ from
the set $\rho^{0}$, $\omega$, $\phi$
is originally formed in the fusion processes with $V$
then converting to the photon.
From isospin invariance the following fusion reactions
giving such a vector meson $V$ are possible
\begin{eqnarray}
&&
(\Pom + f_{2 \Reg}, \rho_{\Reg}) \to \rho^{0}\,, \quad
(\rho_{\Reg}, \Pom + f_{2 \Reg}) \to \rho^{0}\,, \nonumber \\
&&
(\Pom + f_{2 \Reg}, \Ode + \omega_{\Reg}) \to \omega, \phi \,, \quad
(\Ode + \omega_{\Reg}, \Pom + f_{2 \Reg}) \to \omega, \phi \,,\nonumber \\
&&
(a_{2 \Reg}, \Ode + \omega_{\Reg}) \to \rho^{0} \,, \quad
(\Ode + \omega_{\Reg}, a_{2 \Reg}) \to \rho^{0} \,,\nonumber \\
&&
(a_{2 \Reg}, \rho_{\Reg}) \to \omega, \phi \,, \quad
(\rho_{\Reg}, a_{2 \Reg}) \to \omega, \phi \,.
\label{2.15.1}
\end{eqnarray}
Finally the $V \to \gamma$ transition is treated 
in the standard way; see (3.23)--(3.25) of \cite{Ewerz:2013kda}.
Our \textit{Ansatz} for the $\Pom \rho_{\Reg} \rho$ vertex
follows the one for the $\Pom \rho \rho$
in (3.47) of \cite{Ewerz:2013kda}
with the replacements 
$a_{\Pom \rho \rho} \to a_{\Pom \rho_{\Reg} \rho}$,
$b_{\Pom \rho \rho} \to b_{\Pom \rho_{\Reg} \rho}$,
and similarly for the $\Pom \omega_{\Reg} \omega$ vertex
as well as for $f_{2 \Reg}$ in the place of $\Pom$.
All vertices occurring here were discussed in
\cite{Ewerz:2013kda,Bolz:2014mya,Lebiedowicz:2014bea,
Lebiedowicz:2018eui,Lebiedowicz:2019boz} 
except for $\Pom \Ode \omega$, $f_{2 \Reg} \Ode \omega$,
$f_{2 \Reg} \Ode \phi$, and $a_{2 \Reg} \Ode \rho^{0}$.
In a first approximation, at the high energies discussed here,
we shall set these vertices to zero.

As an example we discuss one diagram contributing 
to the $\Reg_{-} - \Pom$ process (\ref{PR}).
The $\Pom \rho_{\Reg}$-exchange amplitude can be written as
\begin{eqnarray}
{\cal M}_{\mu}^{(\Pom \rho_{\Reg})}&=& (-i)\,
\bar{u}(p_{1}',\lambda_{1}) i \Gamma^{(\Pom pp)}_{\alpha_{1} \beta_{1}}(p_{1}',p_{a}) u(p_{a},\lambda_{a})\,
i \Delta^{(\Pom) \,\alpha_{2}\beta_{2},\alpha_{1}\beta_{1}}(s_{1},t_{1})\,
i \Gamma^{(\Pom \rho_{\Reg} \rho)}_{\sigma_{1}\kappa_{1} \alpha_{2}\beta_{2}}(k,q_{2})
\nonumber\\
&&\times 
i \Delta^{(\rho)\, \sigma_{1}\sigma_{2}}(k)\,
i \Gamma_{\sigma_{2} \mu}^{(\rho \to \gamma)}\,
i \Delta^{(\rho_{\Reg}) \,\kappa_{1}\kappa_{2}}(s_{2},t_{2})\,
\bar{u}(p_{2}',\lambda_{2}) i \Gamma^{(\rho_{\Reg} pp)}_{\kappa_{2}}(p_{2}',p_{b}) u(p_{b},\lambda_{b})
\nonumber\\
&=&
ie\frac{m_{\rho}^{2}}{\gamma_{\rho}} \Delta_{T}^{(\rho)}(k^{2})\,
\bar{u}(p_{1}',\lambda_{1}) \Gamma^{(\Pom pp)}_{\alpha \beta}(p_{1}',p_{a}) u(p_{a},\lambda_{a})\,
\frac{1}{2 s_{1}}
\big(-i s_{1} \alpha_{\Pom}'\big)^{\alpha_{\Pom}(t_{1})-1}
\nonumber\\
&&\times 
\left[2 a_{\Pom \rho_{\Reg} \rho} \Gamma_{\mu \kappa}^{(0)\, \alpha \beta}(k,-q_{2})
-b_{\Pom \rho_{\Reg} \rho} \Gamma_{\mu \kappa}^{(2)\, \alpha \beta}(k,-q_{2}) \right]
F_{M}(t_{1})F_{M}(t_{2})F^{(\rho)}(k^{2})
\nonumber\\
&&\times 
\frac{1}{M_{-}^{2}}
\big(-i s_{2} \alpha_{\rho_{\Reg}}'\big)^{\alpha_{\rho_{\Reg}}(t_{2})-1}\,
\bar{u}(p_{2}',\lambda_{2}) \Gamma^{(\rho_{\Reg} pp)\,\kappa}(p_{2}',p_{b}) u(p_{b},\lambda_{b})\,.
\label{PR_amp}
\end{eqnarray}
For real photons ($k^{2} = 0$) we have
$\Delta_{T}^{(\rho)}(k^{2}) = -1/m_{\rho}^{2}$ and $F^{(\rho)}(k^{2}) = 1$.
The $\gamma \rho^{0}$ coupling $\gamma_{\rho}$
is given by (3.23)--(3.25) of \cite{Ewerz:2013kda}.
For the $\Pom \rho_{\Reg} \rho$ coupling
parameters we assume that
$a_{\Pom \rho_{\Reg} \rho} = a_{\Pom \rho \rho}$,
$b_{\Pom \rho_{\Reg} \rho} = b_{\Pom \rho \rho}$
and use the values $a_{\Pom \rho \rho} = 0.45$~GeV$^{-3}$,
$b_{\Pom \rho \rho} = 6.5$~GeV$^{-1}$
from Table~1 of \cite{Bolz:2014mya}.
We have checked that these parameters
give a good description of 
the HERA data \cite{ZEUS:1997rof,H1:2020lzc} 
for the $\gamma p \to \rho^{0} p$ reaction.
The $\rho_{\Reg} \Pom$-exchange amplitude is obtained from
(\ref{PR_amp}) with
the replacements (\ref{replacement}).

In a similar way we obtain the $\Pom \omega_{\Reg}$- 
and $\omega_{\Reg} \Pom$-exchange amplitudes in (\ref{PR}).
Here, in the calculations, we include $\phi$-$\omega$
mixing and we take the coupling parameters
found in (B1), (B4), (B10), and (B11) 
of \cite{Lebiedowicz:2019boz}.

For the amplitude with the odderon exchange,
${\cal M}_{\mu}^{(\Ode-\Pom)}$ (\ref{PO}),
we consider only the two contributions
$(\Ode, \Pom) \to \phi \to \gamma$ 
and $(\Pom, \Ode) \to \phi \to \gamma$.
We use the relations given in Sec.~II~B of 
\cite{Lebiedowicz:2019boz} 
for the $pp \to pp \phi$ reaction. 
Thus, the amplitude for the $\Ode \Pom$-exchange contribution
${\cal M}_{\mu}^{(\Ode \Pom)}$
to $pp \to pp \gamma$
is obtained from (2.26) of \cite{Lebiedowicz:2019boz}
by the replacement
$i \Gamma_{\kappa}^{(\phi KK)}(p_{3},p_{4})
\to 
i \Gamma_{\kappa \mu}^{(\phi \to \gamma)}$.
The same replacement holds for the $\Pom \Ode$-exchange amplitude.
For the $\Ode-\Pom$-fusion processes we shall use 
the double-Regge-pole \textit{Ansatz} \cite{Lebiedowicz:2022nnn} for the odderon.
With this \textit{Ansatz} we were
able to describe the $\rho$ parameter,
the ratio of the real part to the imaginary part of the forward 
$pp$-elastic-scattering amplitude,
as measured by the TOTEM \cite{TOTEM:2017sdy} and ATLAS \cite{ATLAS:2022mgx} Collaborations;
see the discussion in Sec.~IV~A of \cite{Lebiedowicz:2022nnn}.

The quantities needed to calculate 
the fusion amplitudes using the VMD approach
were discussed in
\cite{Ewerz:2013kda,Lebiedowicz:2014bea,Lebiedowicz:2018eui,Lebiedowicz:2019boz}.
In \cite{Ewerz:2013kda} elastic 
and total $\rho^{0} p$ cross sections were discussed.
There, the $\Pom \rho \rho$ 
and $f_{2 \Reg} \rho \rho$ coupling
constants were estimated assuming that at high-energies
the total cross section for transversely polarized 
$\rho^{0}$ mesons
equals to the average of the $\pi^{\pm}p$ cross sections.
In analogy to the $\rho^{0} p$ scattering
the elastic and total cross section 
for $\omega p$ and $\phi p$ were treated in
\cite{Lebiedowicz:2018eui,Lebiedowicz:2019boz}.
From a comparison of our model
for the $\gamma p \to V p$ processes
to the experimental data, especially those from HERA,
the relevant coupling constants and the form-factor parameters
for the pomeron and reggeon exchanges were found.

We use for the purely diffractive fusion contributions
the default values of the reggeon
trajectories from Sec.~3 of \cite{Ewerz:2013kda},
but for the soft pomeron we take in our calculations
\mbox{$\epsilon_{\Pom} = 0.0865$},
as determined in \cite{Lebiedowicz:2022nnn}
by comparison with high-energy $pp$ elastic scattering data,
instead of $\epsilon_{\Pom} = 0.0808$,
the default value from \cite{Ewerz:2013kda};
see the discussion in Appendix~\ref{sec:appendixA}.
We also use the exponential pomeron/reggeon-proton form factor
$F(t) = \exp(-b\,|t|)$ [instead of $F_{1}(t)$ given in 
(3.29) of \cite{Ewerz:2013kda}]
with $b = 2.95$~GeV$^{-2}$
adjusted to the TOTEM data \cite{TOTEM:2017sdy,TOTEM:2018hki}
on $pp$ elastic scattering for $\sqrt{s} = 13$~TeV
(see Fig.~5 of \cite{Lebiedowicz:2022nnn}).

\section{Results and discussions}
\label{sec:3}

We start by showing our results for the $pp \to pp \gamma$ reaction
from the photon-$\Pom/\Reg$-fusion processes.
Calculations were done for $\sqrt{s} = 13$~TeV,
$0.1~{\rm GeV} < k_{\perp} < 1$~GeV,
and for photon rapidities $|\rm y| < 2.5$.
In this kinematic range
the soft-pomeron term ($\Pom_{1}$) gives the dominant contribution
while the hard-pomeron term ($\Pom_{0}$) and 
the reggeon term are negligibly small.
There also the FIT~1 (\ref{FIT1}) and FIT~2 (\ref{FIT2}) 
parametrizations for the $\Pom_{j} \gamma^{*} \gamma^{*}$ and
$\Reg_{+}\gamma^{*} \gamma^{*}$ coupling functions hardly differ.
In the calculations for the photoproduction contribution
we use the FIT~2 parametrization. 

\begin{figure}[!ht]
\includegraphics[width=0.49\textwidth]{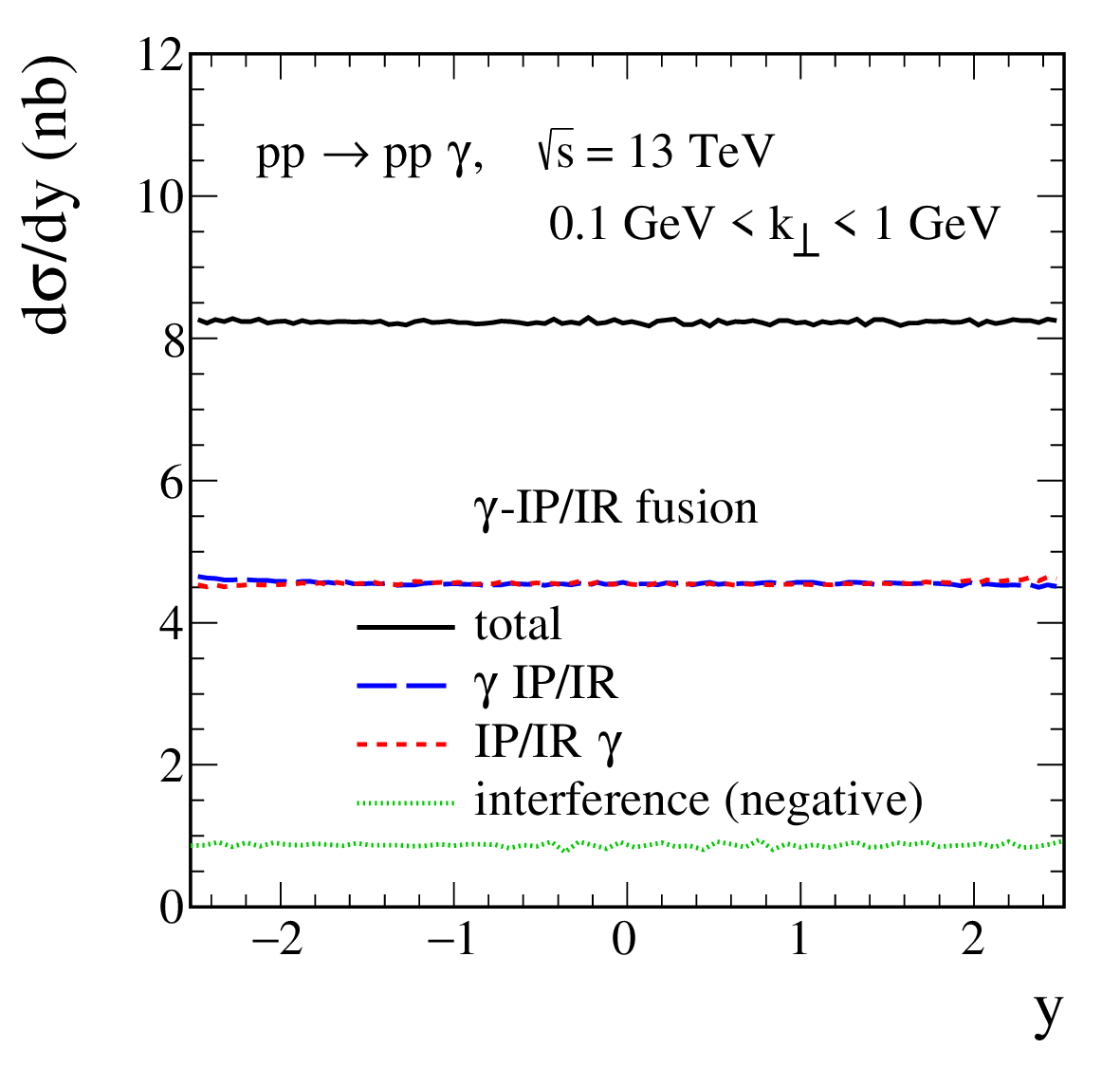}
\includegraphics[width=0.49\textwidth]{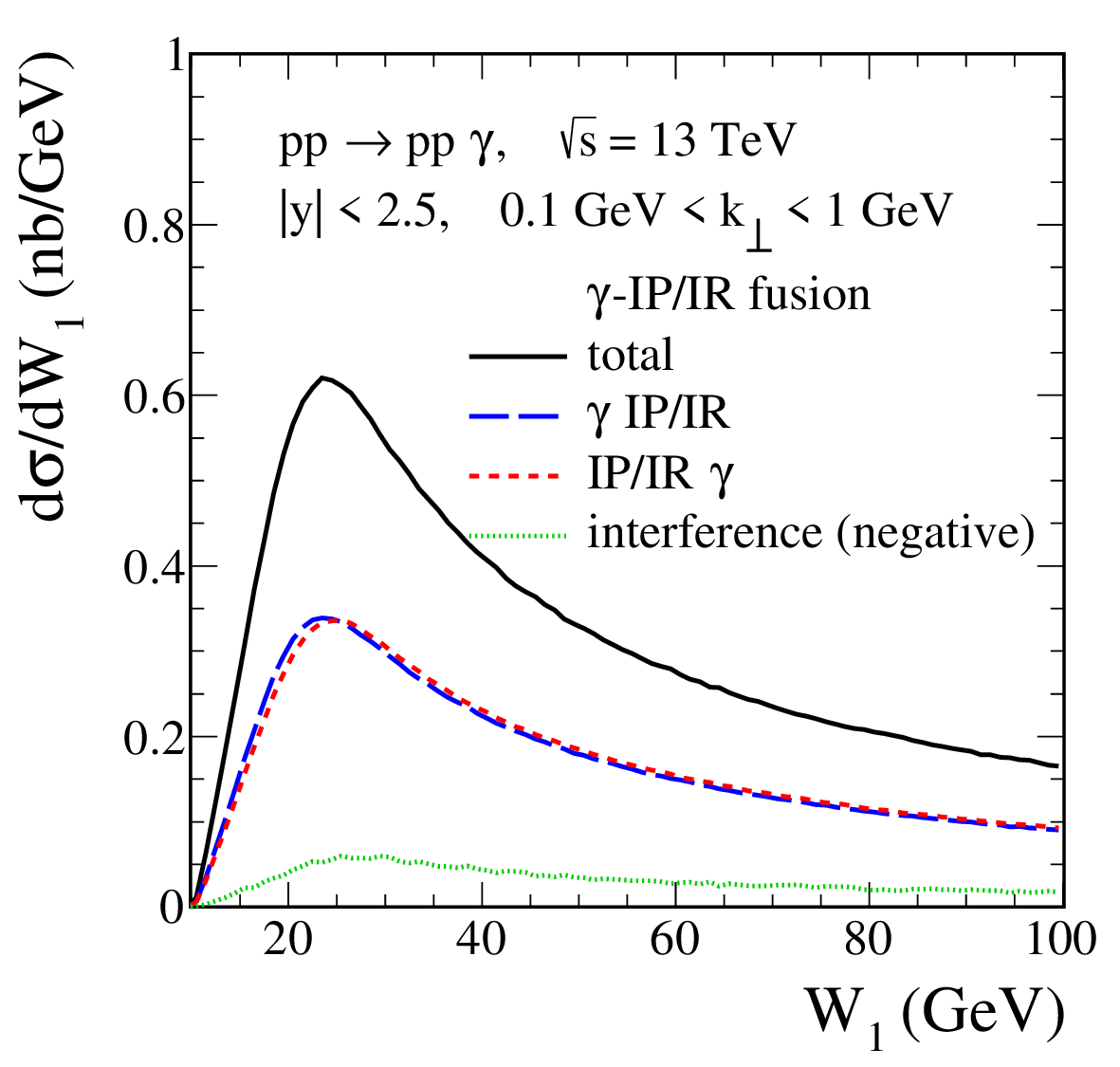}
\caption{\label{fig:1}
\small
The differential distributions 
in the rapidity of the photon and
in the subenergy $W_{1}$ of the $\gamma p$ system
for the $pp \to pp \gamma$ reaction
via the $\gamma - \Pom/\Reg$-fusion processes.
The calculations were done for $\sqrt{s} = 13$~TeV and
with cuts on $|{\rm y}| < 2.5$ and 
$0.1~{\rm GeV} < k_{\perp} < 1$~GeV.
The solid line corresponds to the complete result (total),
the blue long-dashed and red dashed lines
correspond to the $\gamma \Pom/\Reg$ contribution
[see Fig.~\ref{fig:pp_pp_gam_CEP_photoprod}~(a)] and 
the $\Pom/\Reg \gamma$ contribution
[see Fig.~\ref{fig:pp_pp_gam_CEP_photoprod}~(b)], respectively.
The destructive interference term 
is shown separately by the green dotted line.}
\end{figure}

In Fig.~\ref{fig:1} we show the distributions
in ${\rm y}$, rapidity of the photon,
and in $W_{1}$, the subenergy of the $\gamma p$ system.
We present the complete result (denoted by ``total'')
and the results corresponding
to the diagrams shown in 
Fig.~\ref{fig:pp_pp_gam_CEP_photoprod}(a) and 
Fig.~\ref{fig:pp_pp_gam_CEP_photoprod}(b) separately.
The interference term between
the $\gamma \Pom/\Reg$ contribution [diagram (a)] and
the $\Pom/\Reg \gamma$ contribution [diagram (b)]
is also shown.
This interference effect is destructive.

\begin{figure}[!ht]
\includegraphics[width=0.49\textwidth]{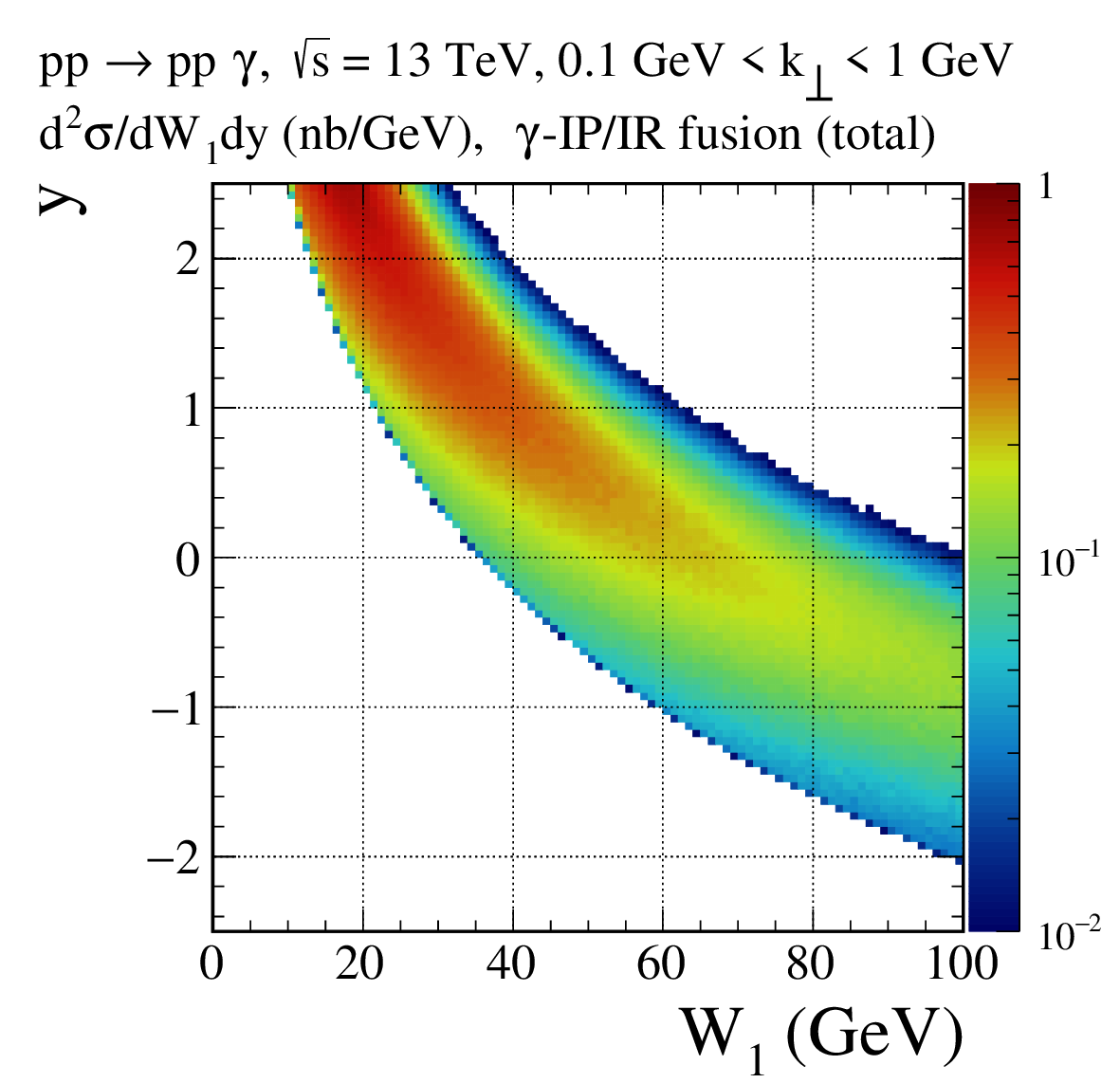}
\includegraphics[width=0.49\textwidth]{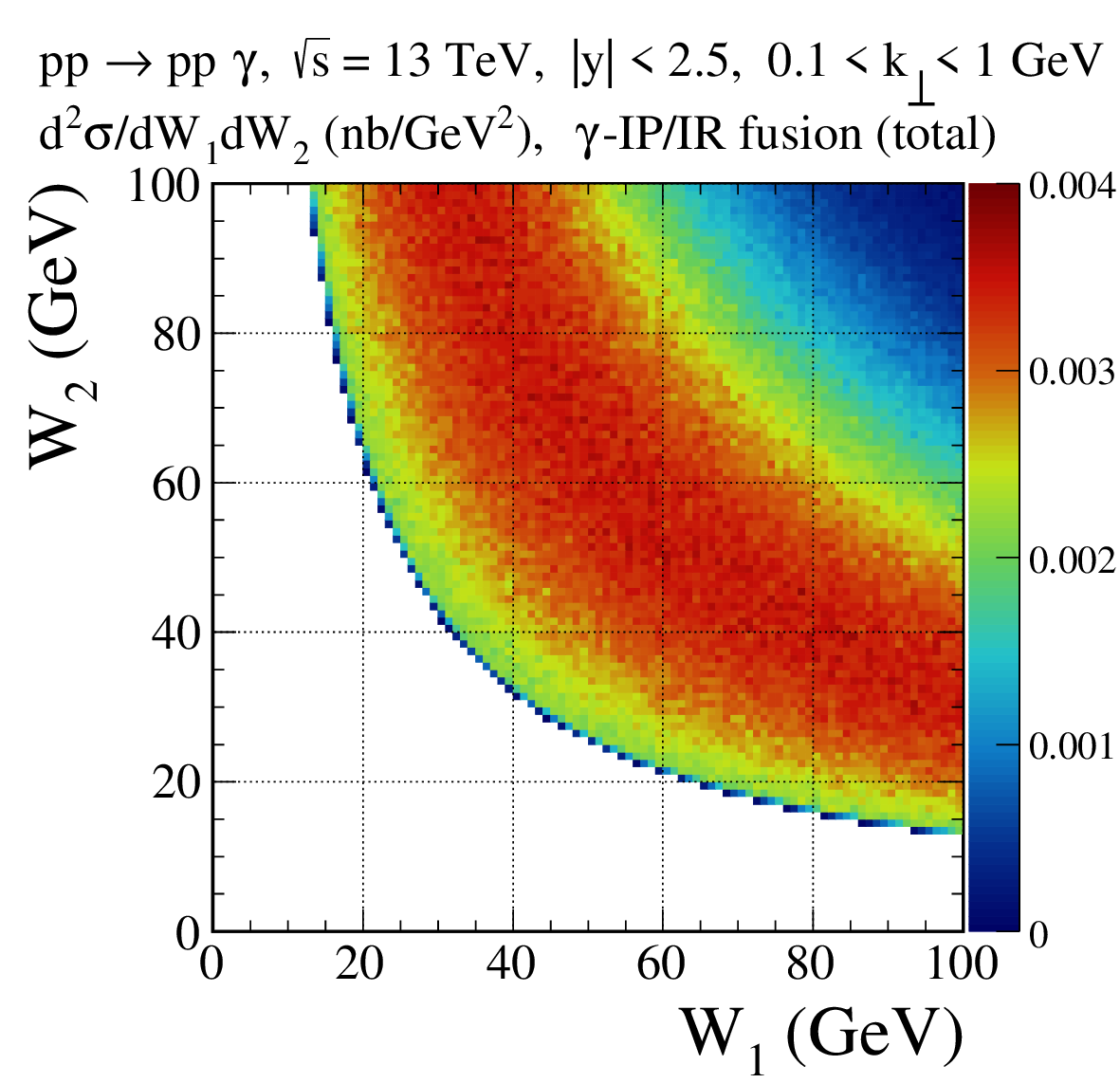}
\caption{\label{fig:2}
\small
The two-dimensional distributions in
$(W_{1}, {\rm y})$ and $(W_{1}, W_{2})$
for the $pp \to pp \gamma$ reaction
via the $\gamma - \Pom/\Reg$-fusion processes.
The calculations were done for $\sqrt{s} = 13$~TeV and
with cuts on $|{\rm y}| < 2.5$ and 
$0.1~{\rm GeV} < k_{\perp} < 1$~GeV.}
\end{figure}

Figure~\ref{fig:2} shows that limiting ourselves to $|\rm y| < 2.5$
and $0.1~{\rm GeV} < k_{\perp} < 1$~GeV we avoid
the low $W_{1}$ and low $W_{2}$ regions
which are a bit less under theoretical control in our model;
see the discussion in Sec.~\ref{sec:2B}.
The left panel shows the distribution in $(W_{1}, {\rm y})$.
The cut at $|\rm y| < 2.5$ eliminates small subenergies $W_{1}$
[see (\ref{2.17})] and we have $W_{1} \geqslant 10$~GeV.
For symmetry reasons the $(W_{2}, {\rm y})$ distribution
is obtained by the replacement
$(W_{1}, {\rm y}) \to (W_{2}, -{\rm y})$.
Thus, we also have $W_{2} \geqslant 10$~GeV.
In the right panel of Figure~\ref{fig:2}
we show the $(W_{1}, W_{2})$ distribution which again shows
very clearly that with our cuts we avoid the regions
of small $W_{1}$ and/or $W_{2}$.

We also find small enhancements at photon rapidities 
$|\rm y| \simeq 2.5$, which correspond to low-$W_{1,2}$ regions
(see Fig.~\ref{fig:2}).
This effect is due to the reggeon component and its constructive interference with
the soft-pomeron;\footnote{Note 
that another constructive interference effect,
namely that between the soft- and hard-pomeron components,
plays an important role in the description of HERA DVCS data
in the process $\gamma^{*}(Q^{2}) p \to \gamma p$,
especially for large photon virtualities $Q^{2}$;
see e.g. Fig.~4 of \cite{Lebiedowicz:2022xgi}.
The reggeon contribution is negligibly small there.}
see the left panel of Fig.~\ref{fig:1}.
This effect is more visible in Fig.~\ref{fig:aux} below
for photoproduction at $|\rm y| \simeq 4$.

\begin{figure}[!ht]
\includegraphics[width=0.49\textwidth]{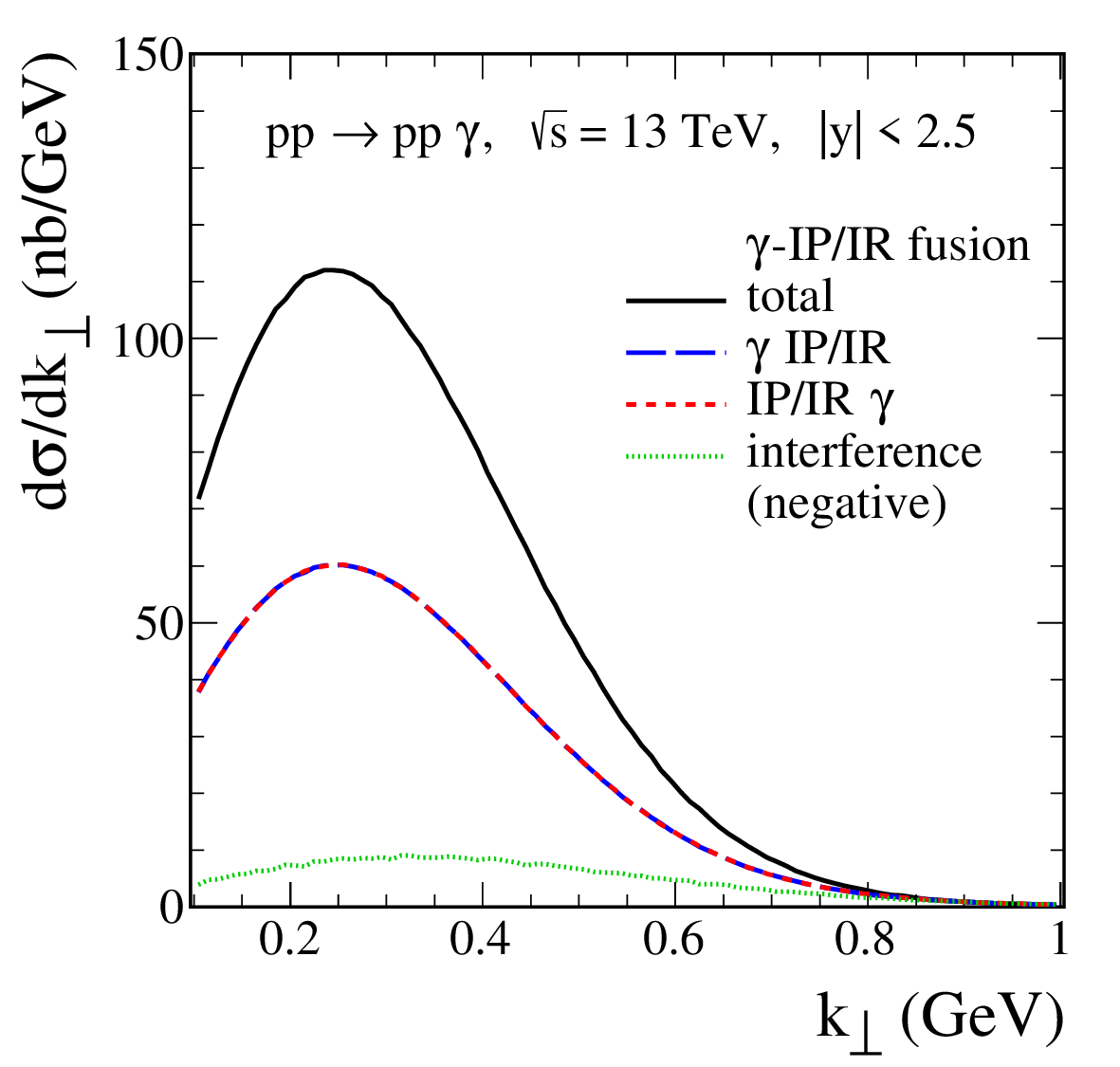}
\includegraphics[width=0.49\textwidth]{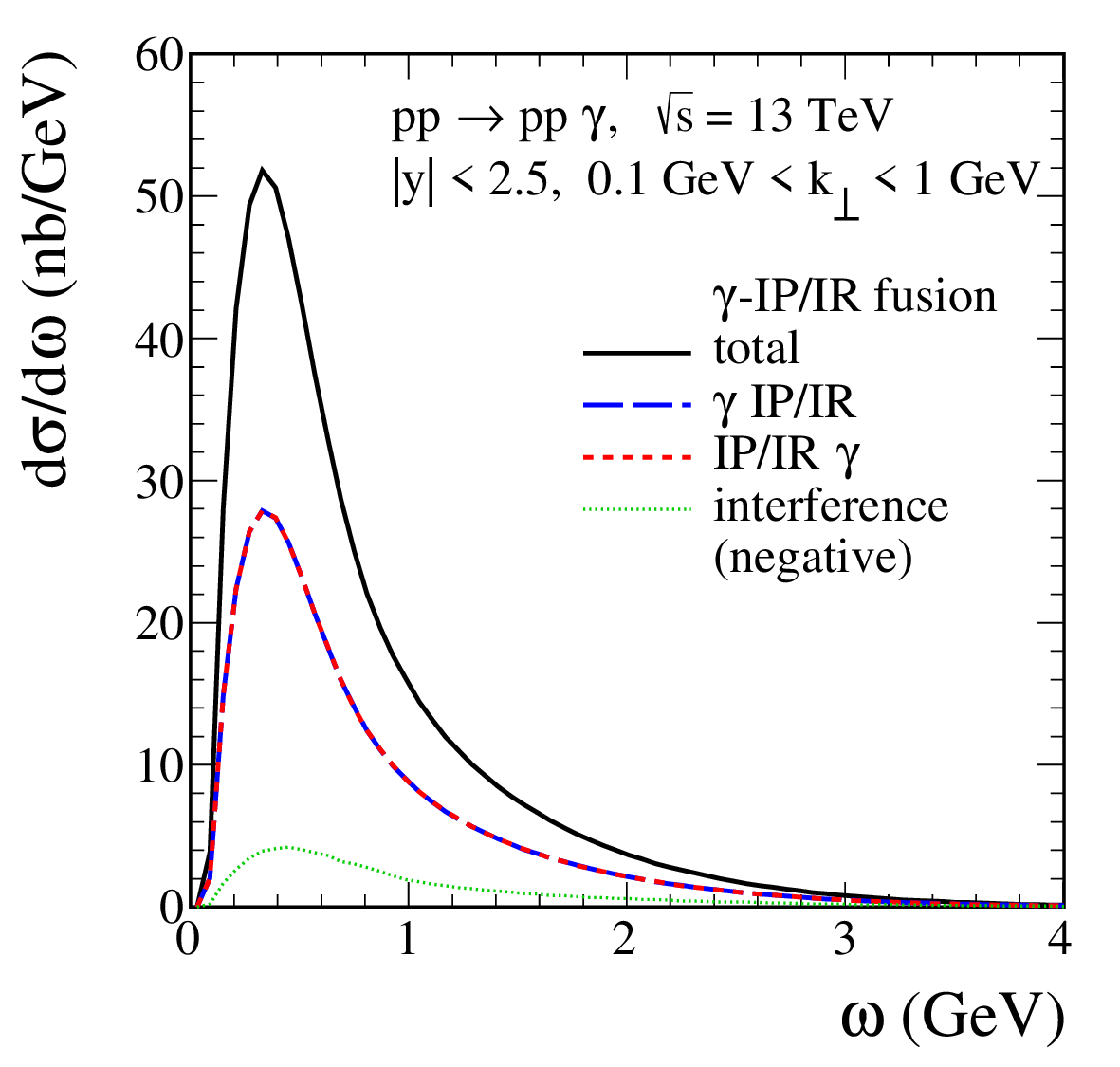}
\caption{\label{fig:3}
\small
The differential distributions 
in transverse momentum of the photon and
in the energy of the photon
for the $pp \to pp \gamma$ reaction
via the $\gamma - \Pom/\Reg$-fusion processes.
The calculations were done for $\sqrt{s} = 13$~TeV and
with cuts on $|{\rm y}| < 2.5$ and 
$0.1~{\rm GeV} < k_{\perp} < 1$~GeV.
The meaning of the lines is the same as in Fig.~\ref{fig:1}.}
\end{figure}
In Fig.~\ref{fig:3} we present the distributions
in $k_{\perp}$ and $\omega$. 
Again we show the complete result (total),
the $\gamma \Pom/\Reg$ and $\Pom/\Reg \gamma$ terms,
and the interference term between them.
The cross sections $d\sigma/dk_{\perp}$ and $d\sigma/d\omega$
gradually increase and reach maxima 
at $k_{\perp} \simeq 0.25$~GeV and
at $\omega \simeq 0.4$~GeV, respectively.
After that both distributions decrease quickly
with increasing $k_{\perp}$ and $\omega$.

\begin{figure}[!ht]
\includegraphics[width=0.49\textwidth]{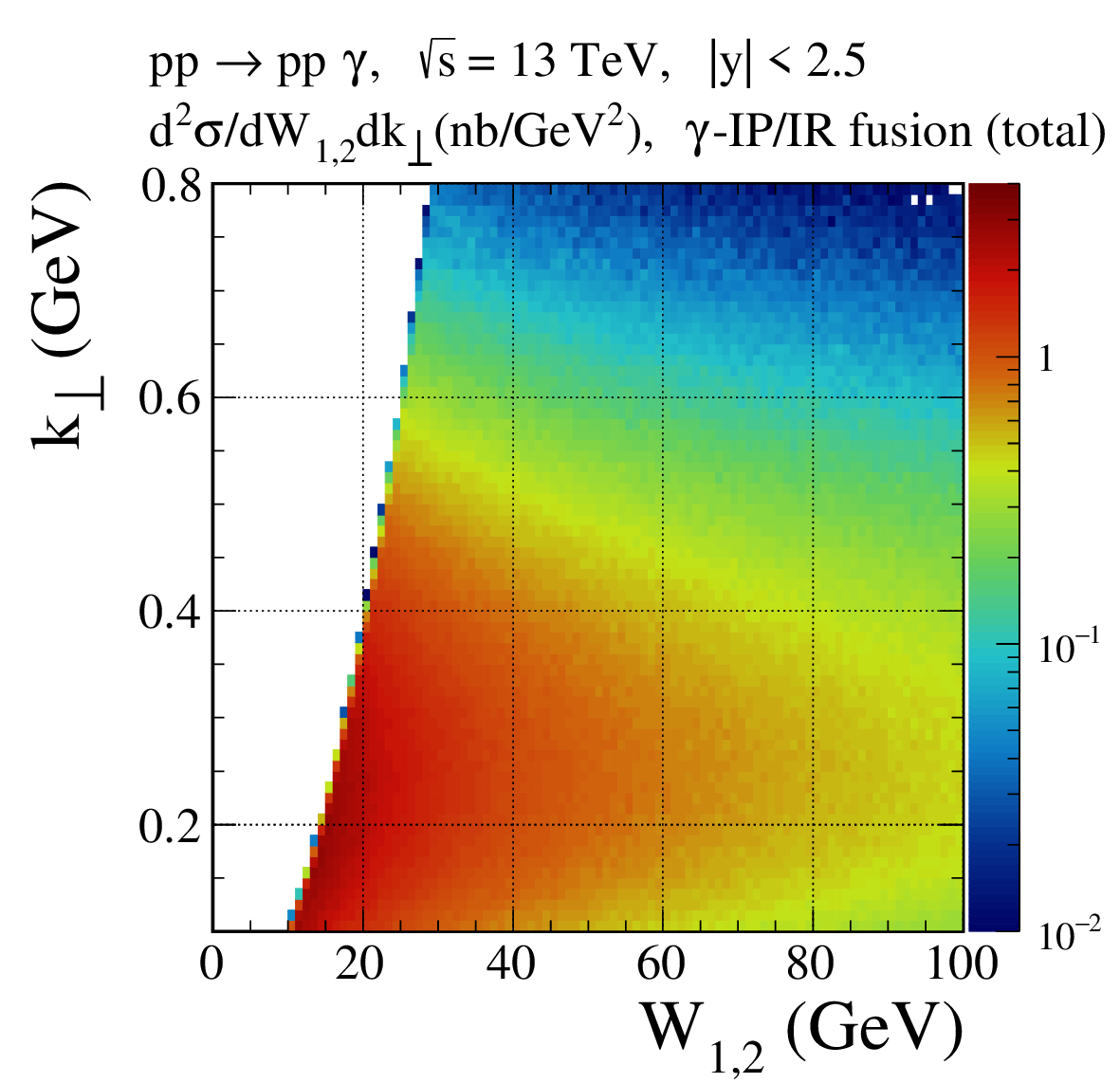}
\includegraphics[width=0.49\textwidth]{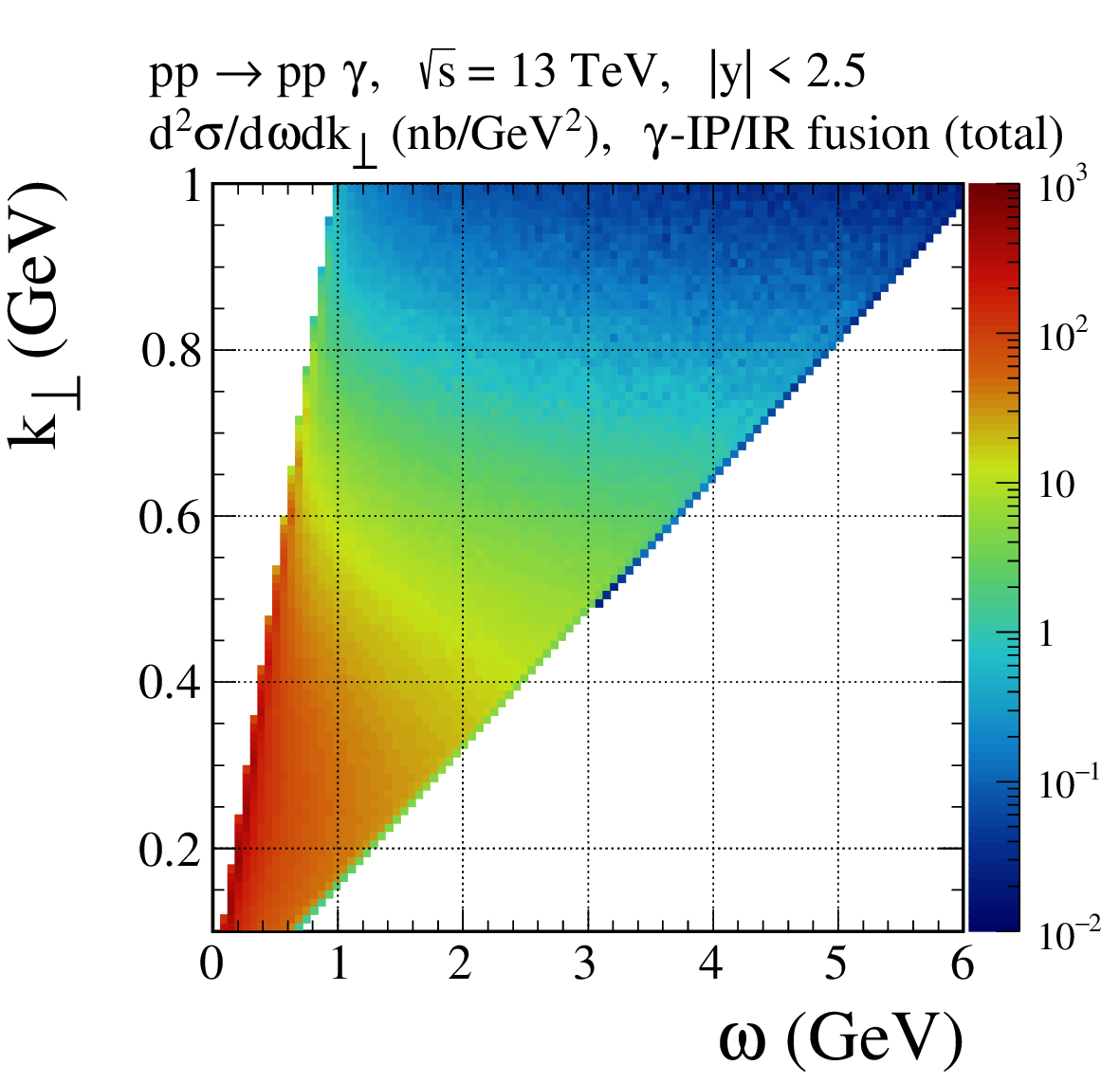}
\caption{\label{fig:4}
\small
The two-dimensional distributions in $(W_{1,2}, k_{\perp})$
and $(\omega, k_{\perp})$ for the $pp \to pp \gamma$ reaction
via the $\gamma - \Pom/\Reg$-fusion processes.
The calculations were done for $\sqrt{s} = 13$~TeV 
and $|{\rm y}| < 2.5$.}
\end{figure}
In Fig.~\ref{fig:4} we show the two-dimensional
differential cross sections in the $W_{1,2}$-$k_{\perp}$ plane
(the left panel) and 
in the $\omega$-$k_{\perp}$ plane (the right panel) 
calculated for $|{\rm y}| < 2.5$.
In the right panel, large $|{\rm y}|$ is near the $\omega$ axis,
and ${\rm y} = 0$ corresponds to $\omega = k_{\perp}$.
The phase-space region $\omega < k_{\perp}$ is forbidden.

\begin{figure}[!ht]
(a)\includegraphics[width=0.46\textwidth]{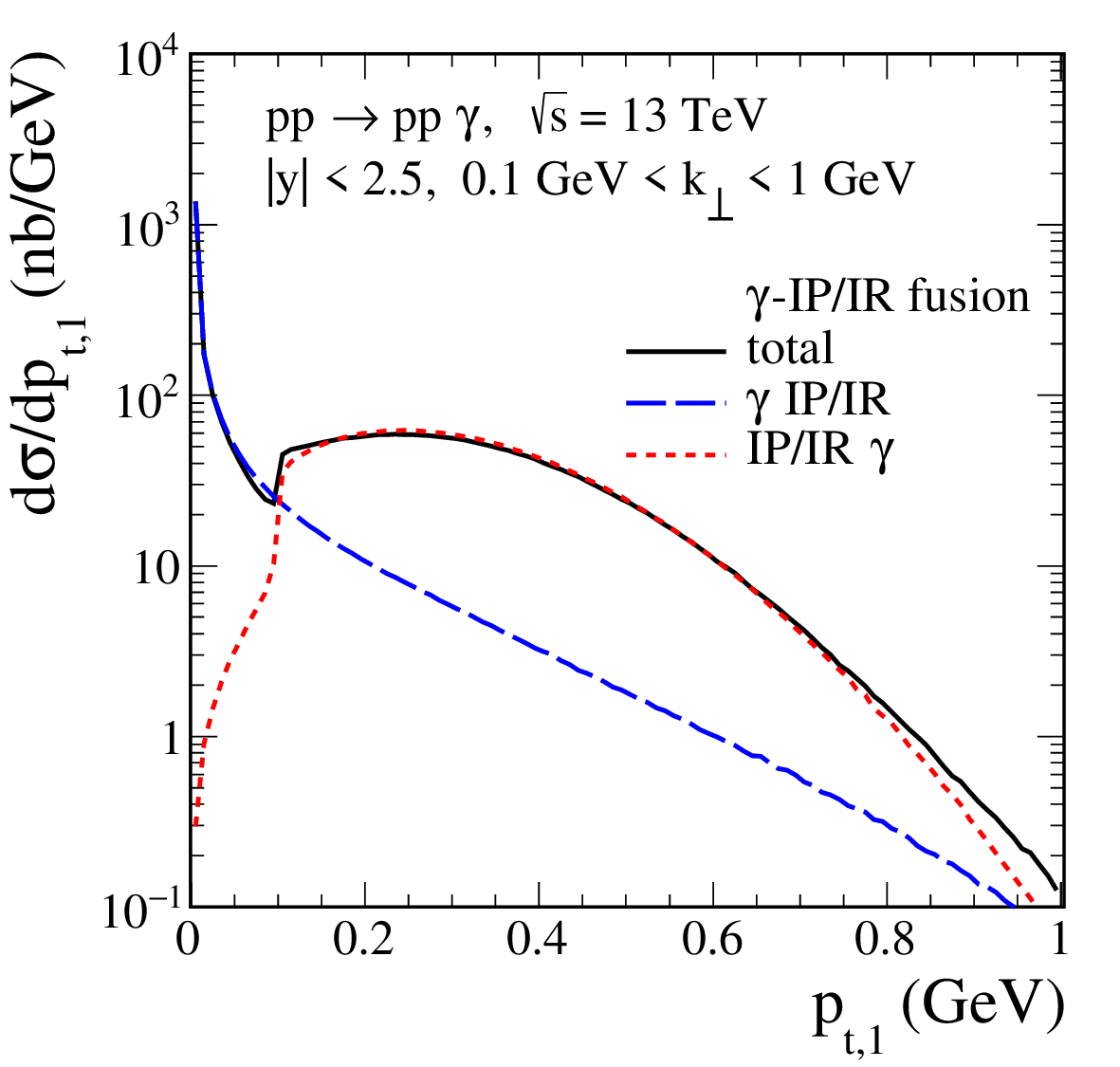}
(b)\includegraphics[width=0.46\textwidth]{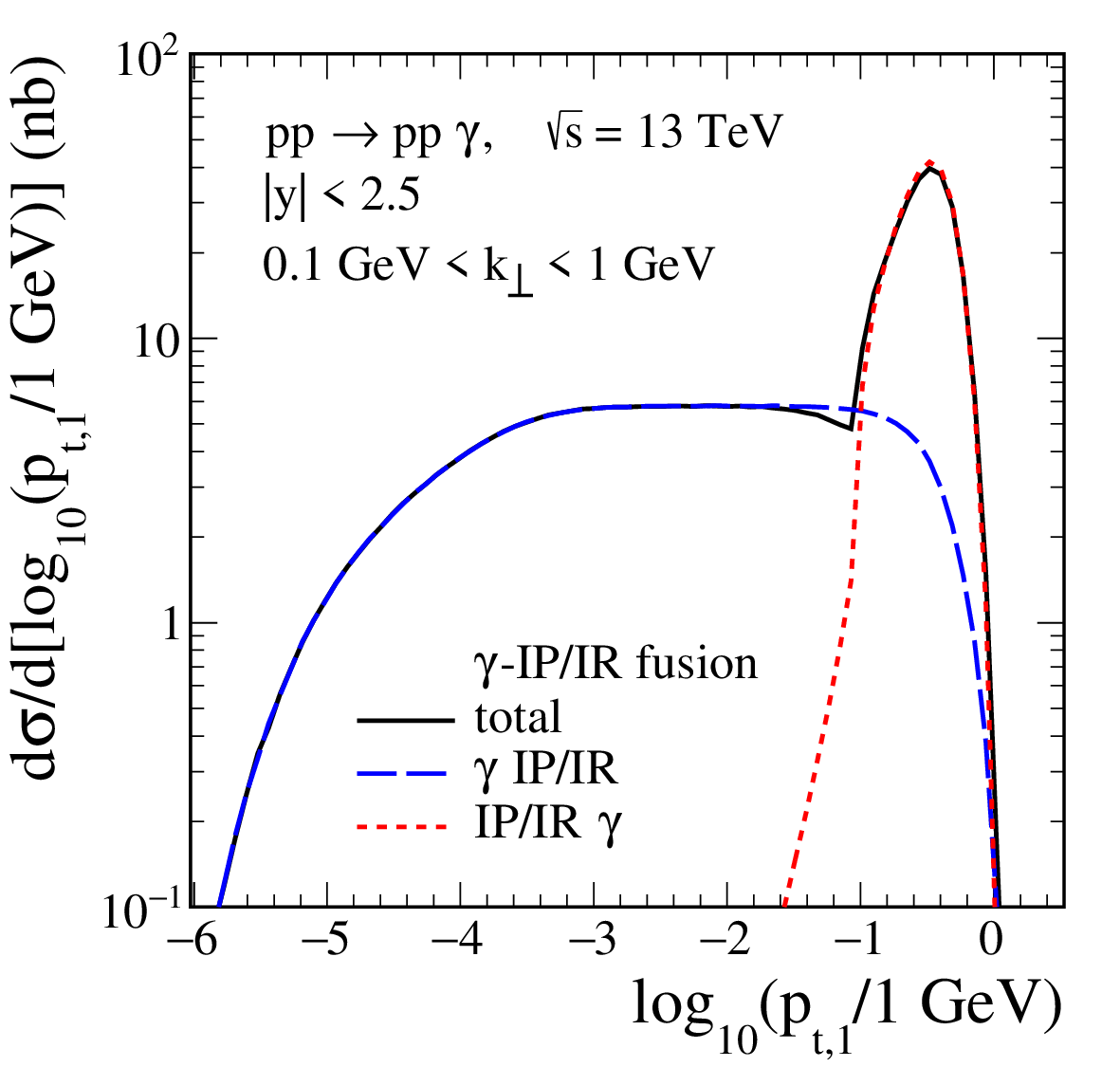}
(c)\includegraphics[width=0.46\textwidth]{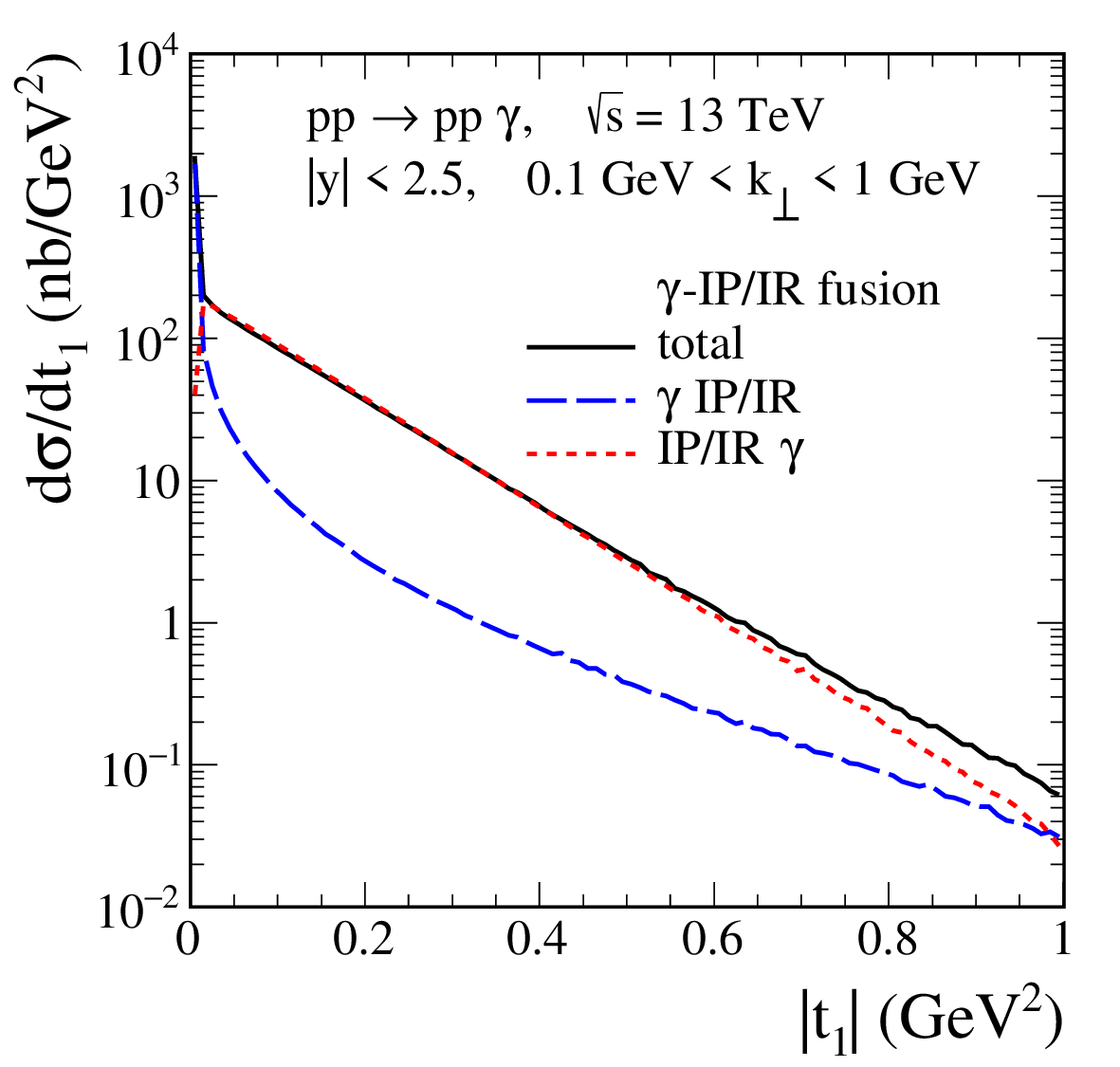}
(d)\includegraphics[width=0.46\textwidth]{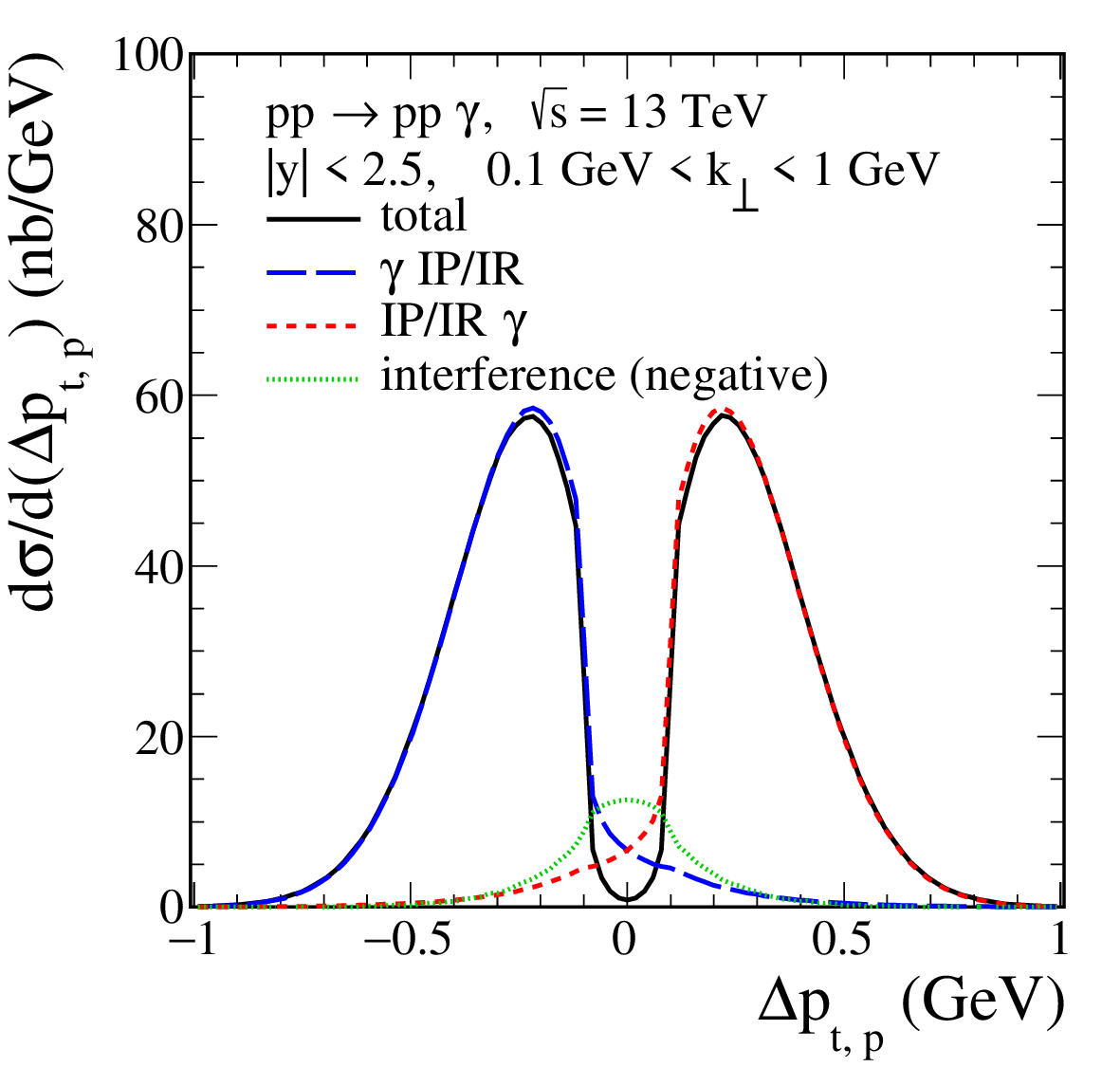}
\caption{\label{fig:5}
\small
The differential distributions 
in transverse momentum of the proton $p(p_{1}')$,
in four-momentum transfer squared $|t_{1}|$,
and in $\Delta p_{t,p} = |\bptap| - |\bptbp|$
for the $pp \to pp \gamma$ reaction
via the $\gamma - \Pom/\Reg$-fusion processes.
The calculations were done for $\sqrt{s} = 13$~TeV and
with cuts on $|{\rm y}| < 2.5$ and 
$0.1~{\rm GeV} < k_{\perp} < 1$~GeV.
The meaning of the lines is the same as in
Fig.~\ref{fig:1}.}
\end{figure}
In Fig.~\ref{fig:5} we show the distributions
in transverse momentum of the proton $p(p_{1}')$
(here $p_{t ,1} = |\bptap|$),
in four-momentum transfer squared ($t_{1}$),
and in $\Delta p_{t,p}$ defined as
\begin{eqnarray}
\Delta p_{t,p} = |\bptap| - |\bptbp|\,.
\label{3.1}
\end{eqnarray}
Figures~\ref{fig:5}(a) and \ref{fig:5}(b) show that 
the low-$p_{t,1}$ region
is dominated by the photon exchange 
in the $\gamma \Pom/\Reg$ term.
This is caused by
the factor $1/t_{1}$ for $\gamma \Pom/\Reg$ from
the photon propagator.
In Fig.~\ref{fig:5}(c), the region of intermediate $|t_{1}|$,
\mbox{$0.01\; {\rm GeV}^{2} < |t_{1}| < 1\; {\rm GeV}^{2}$},
is governed by the pomeron exchange
in the $\Pom/\Reg \gamma$ term.
Not shown is the large-$|t_{1}|$ region, 
$|t_{1}| > 1$~GeV$^{2}$.
We find that there again the $\gamma \Pom/\Reg$ term dominates.
This is due to the $1/|t_{1}|$ fall-off of the photon
propagator in the $\gamma \Pom/\Reg$ term winning over
the exponential fall-off in $|t_{1}|$ of the pomeron part
in the $\Pom/\Reg \gamma$ term.
Note that for the CEP of a photon we find
$|t_{1}|$ and $p_{t,1}$ distributions which are strongly peaked
at very small $|t_{1}|$ and $p_{t,1}$, respectively.
In contrast, for the diffractive bremsstrahlung mechanism
photons hardly contribute to very small values of $p_{t,p}$
and $d\sigma/dp_{t,p}$ reaches a maximum at
$p_{t,p} \sim \sqrt{|t_{1,2}|} \sim 0.3$~GeV;
see Fig.~7 of \cite{Lebiedowicz:2022nnn}.
The $\Delta p_{t,p}$ distribution
shown in Fig.~\ref{fig:5}(d)
is interesting as there the destructive interference
between the two terms
$\gamma \Pom/\Reg$ and $\Pom/\Reg \gamma$
is sizeable around $\Delta p_{t,p} = 0$. 

\begin{figure}[!ht]
(a)\includegraphics[width=0.46\textwidth]{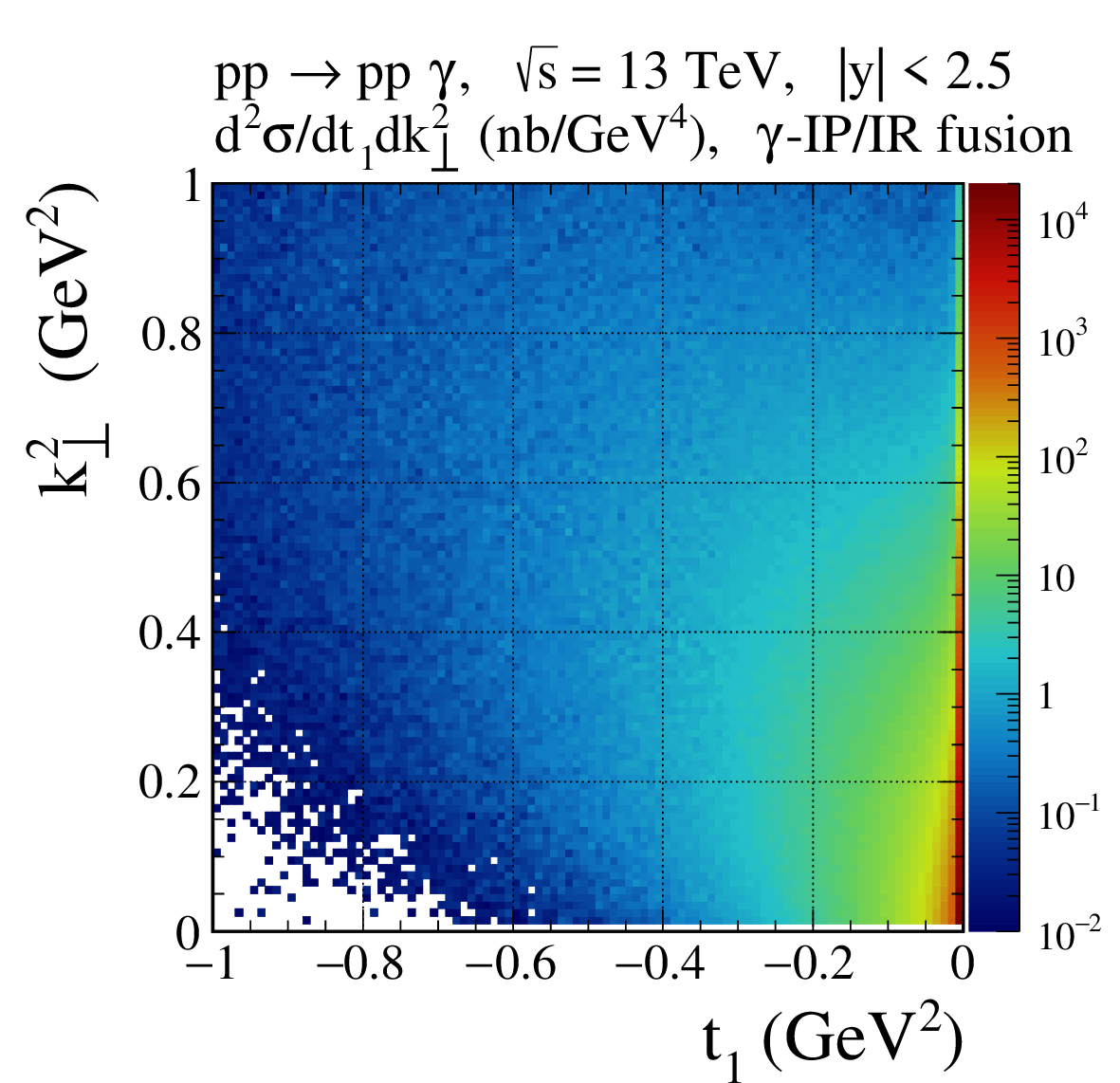}
(b)\includegraphics[width=0.46\textwidth]{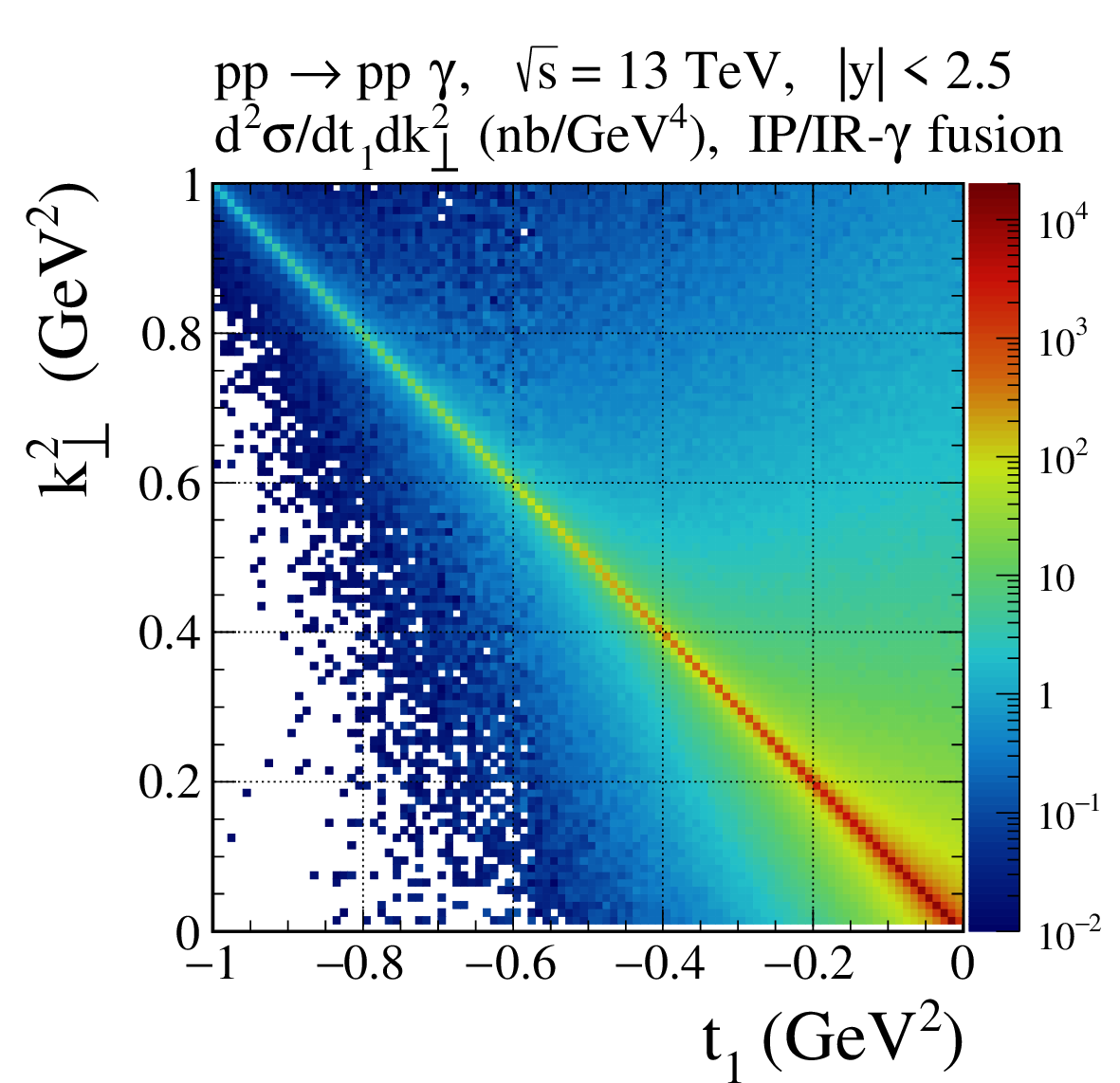}\\
(c)\includegraphics[width=0.46\textwidth]{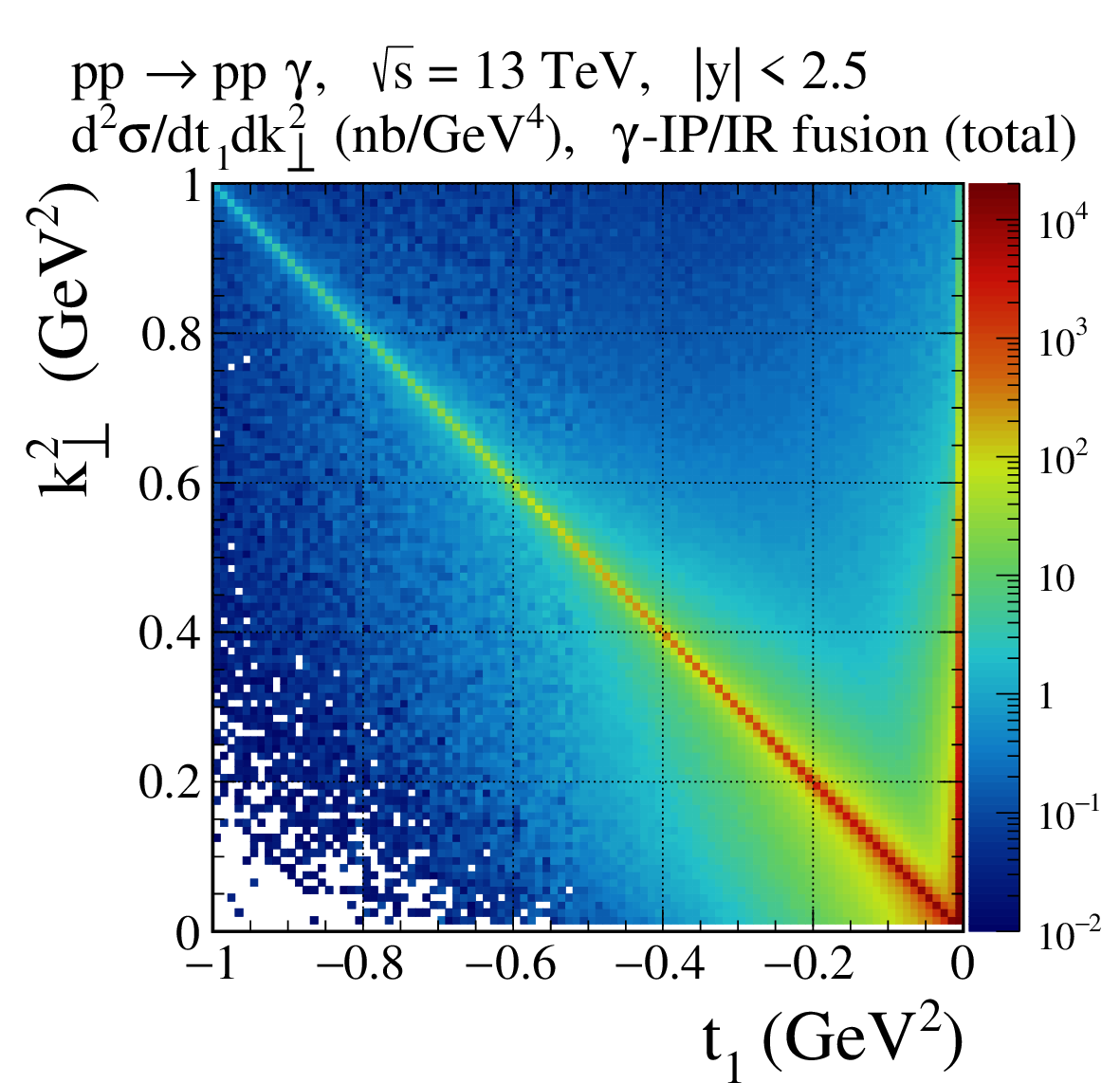}
\caption{\label{fig:6}
\small
The two-dimensional distributions in
$(t_{1}, k_{\perp}^{2})$
for the $pp \to pp \gamma$ reaction.
The calculations were done for $\sqrt{s} = 13$~TeV and
with cuts on $|{\rm y}| < 2.5$ 
and $0.1~{\rm GeV} < k_{\perp} < 1$~GeV.
The top left panel shows the result for only 
the diagram~(a) of Fig.~\ref{fig:pp_pp_gam_CEP_photoprod}
and the top right panel for the diagram~(b) 
of Fig.~\ref{fig:pp_pp_gam_CEP_photoprod}.
The bottom panel shows the complete result
for the $\gamma - \Pom/\Reg$-fusion processes.}
\end{figure}

\begin{figure}[!ht]
\includegraphics[width=0.47\textwidth]{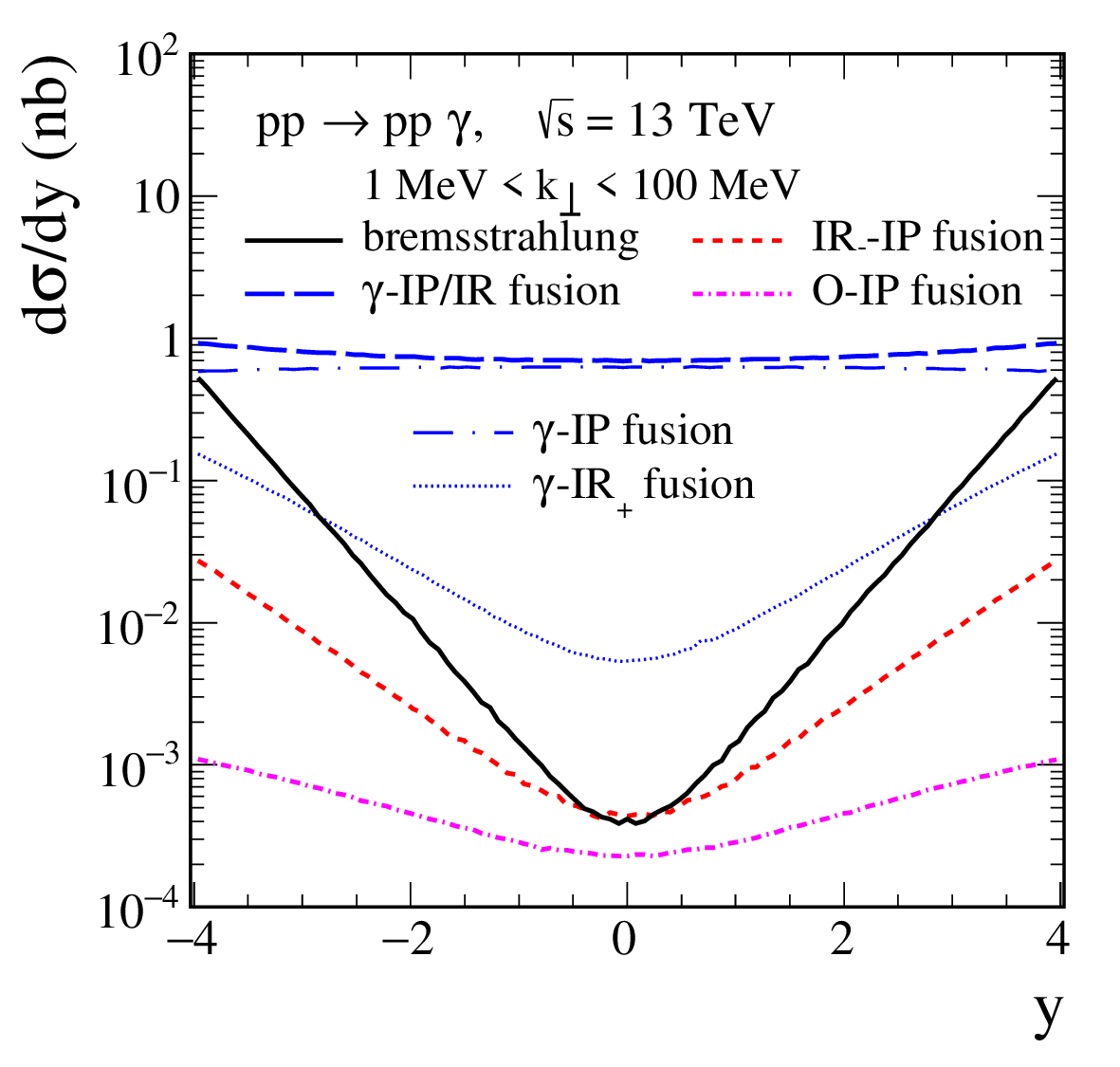}
\includegraphics[width=0.47\textwidth]{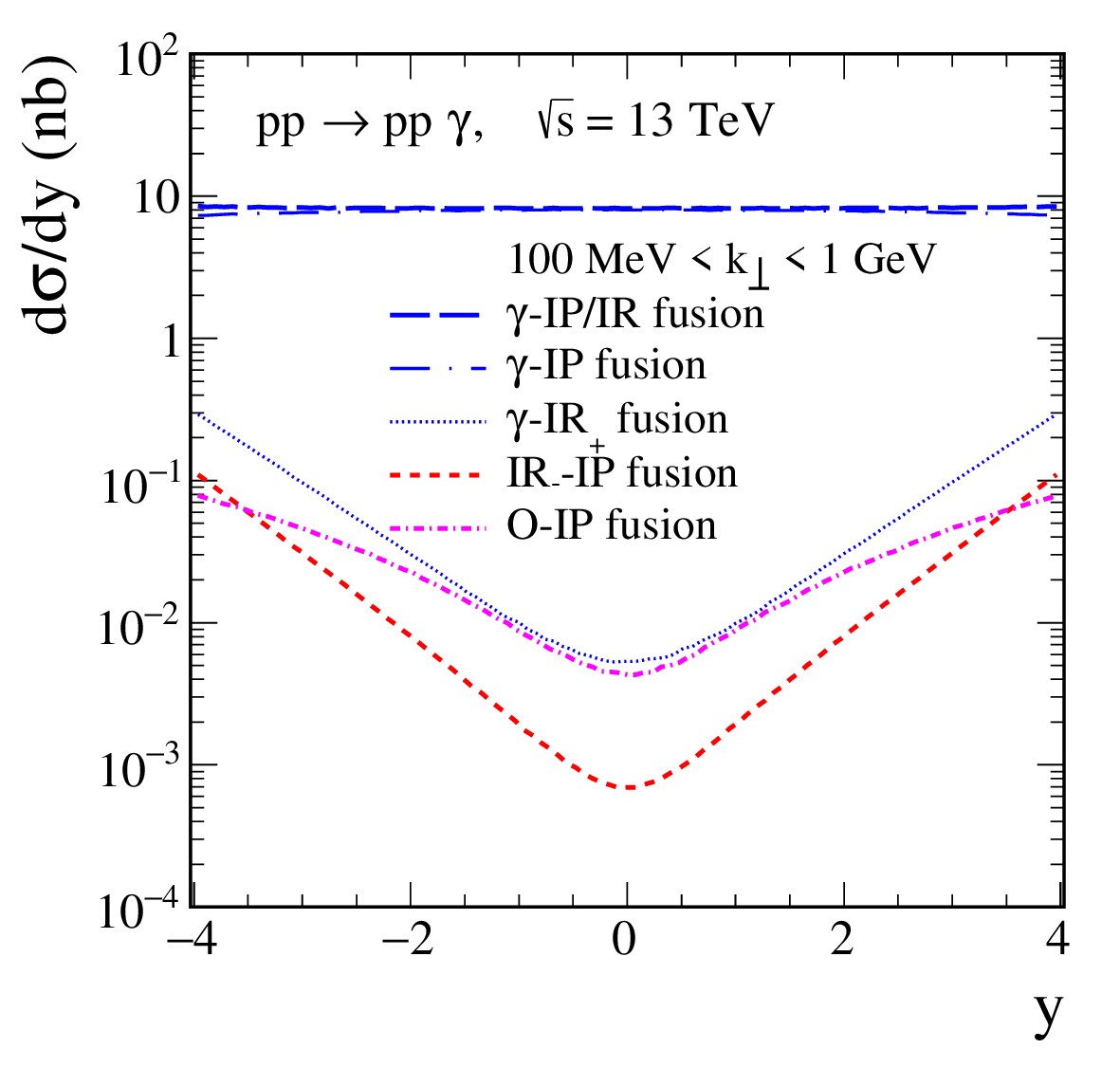}
\includegraphics[width=0.47\textwidth]{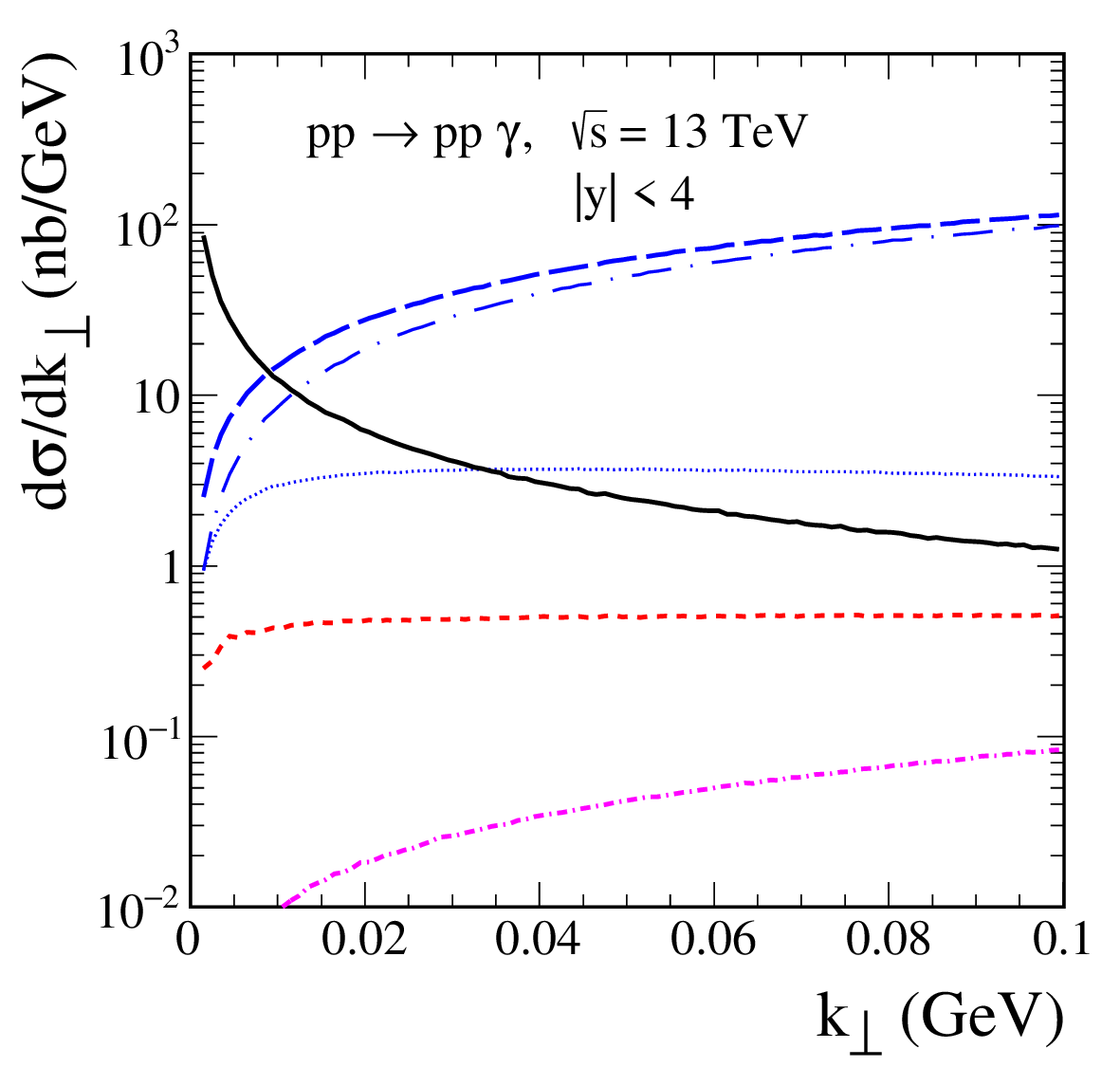}
\includegraphics[width=0.47\textwidth]{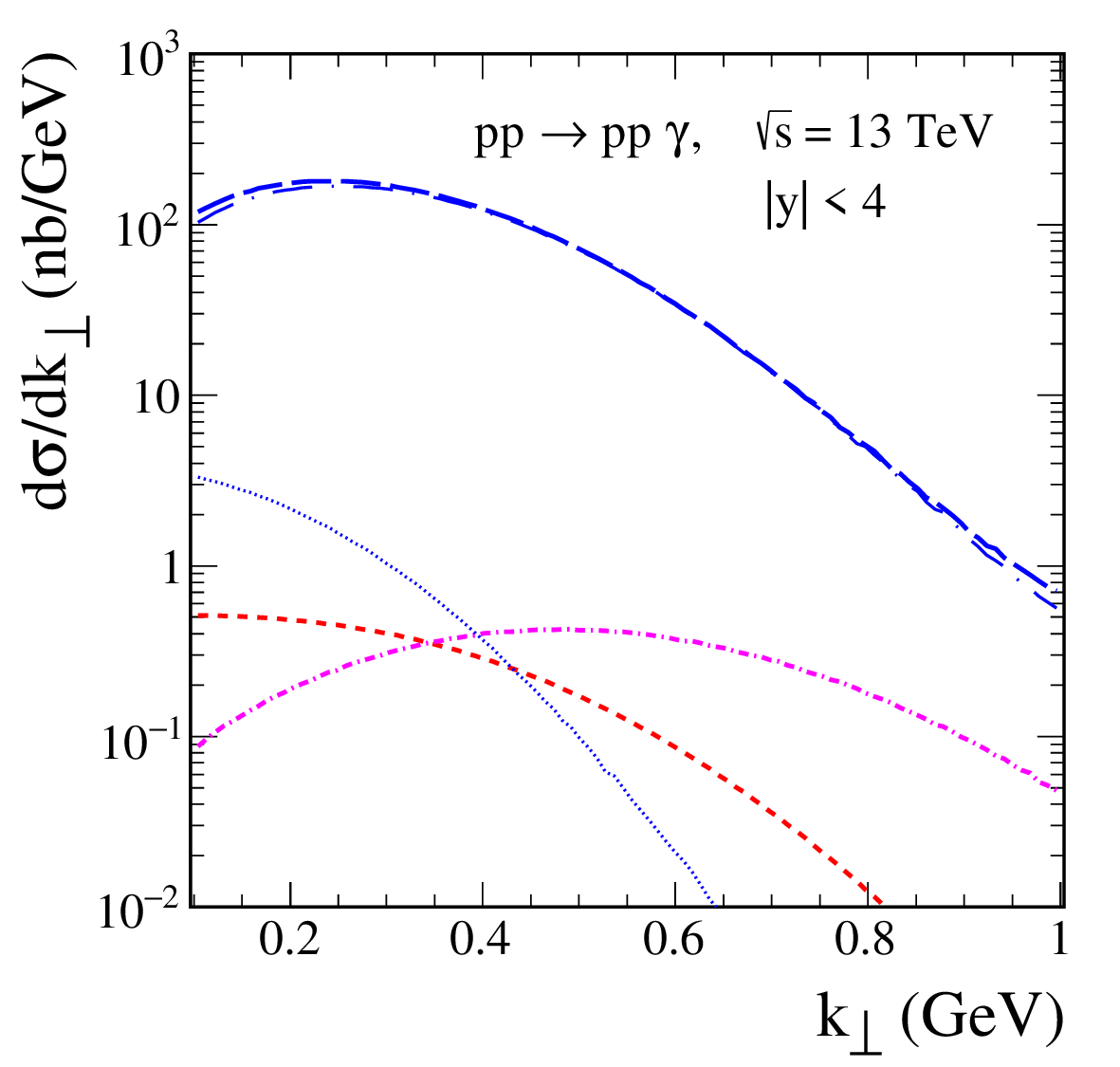}
\caption{\label{fig:aux}
\small
The differential distributions 
for the $pp \to pp \gamma$ reaction
calculated for $\sqrt{s} = 13$~TeV,
$|{\rm y}| < 4$,
\mbox{$1\; {\rm MeV} < k_{\perp} < 100\; {\rm MeV}$}
(left panels)
and 
\mbox{$100\; {\rm MeV} < k_{\perp} < 1\; {\rm GeV}$}
(right panels).
The black solid line corresponds to the diffractive bremsstrahlung,
the blue long-dash-dotted line to the $\gamma-\Pom$-fusion processes
($\gamma \Pom + \Pom \gamma$),
the blue dotted line to the $\gamma-\Reg_{+}$-fusion processes 
($\gamma \Reg_{+} + \Reg_{+} \gamma$), and
the blue long-dashed line to the coherent sum of all photoproduction contributions.
The red dashed line corresponds to the $\Reg_{-} - \Pom$-fusion processes
while the magenta dash-dotted line to the $\Ode - \Pom$ contribution.}
\end{figure}

In Fig.~\ref{fig:6} we show the two-dimensional
distributions in $(t_{1}, k_{\perp}^{2})$.
The results for the two diagrams 
shown in Fig.~\ref{fig:pp_pp_gam_CEP_photoprod} 
and for their coherent sum (total) are presented.
From transverse momentum conservation
the transverse momentum of the final state photon
is the vector sum of the transverse momenta
of virtual photon and pomeron/reggeon.
The plots of Fig.~\ref{fig:6} are easily understood.
For the $\gamma \Pom/\Reg$ term the photon propagator
in Fig.~\ref{fig:pp_pp_gam_CEP_photoprod}(a)
forces the proton $p(p_{1}')$ to go out with very small
transverse momentum, $|\bptap|^{2} \approx - t_{1} \approx 0$.
Then the proton $p(p_{2}')$ will have
$|\bptbp|^{2} \approx |\bk_{\perp}|^{2}$ and this ranges
up to $|\bk_{\perp}|^{2} \approx 1$~GeV$^{2}$;
see Fig.~\ref{fig:6}(a).
For the $\Pom/\Reg \gamma$ term we will correspondingly
have $|\bptbp|^{2} \approx 0$ and
$|\bptap|^{2} \approx - t_{1} \approx |\bk_{\perp}|^{2}$.
This is indeed what is seen in Fig.~\ref{fig:6}(b).
In Fig.~\ref{fig:6}(c) we see the combination of these two effects.

Finally, we compare the results for the $pp \to pp \gamma$ reaction
from the fusion processes (CEP) discussed in this article with
the results corresponding to the diffractive bremsstrahlung 
discussed in \cite{Lebiedowicz:2022nnn}.
We consider separately the bremsstrahlung,
the photoproduction $\gamma - \Pom/\Reg$, 
and the two diffractive contributions 
$\Reg_{-} - \Pom$ and $\Ode - \Pom$.
We denote, for brevity, the coherent sum of the contributions
$\gamma \Pom$ and $\Pom \gamma$ by $\gamma - \Pom$,
the coherent sum of $\gamma \Reg_{+}$ and $\Reg_{+} \gamma$ 
by $\gamma - \Reg_{+}$,
and the complete photoproduction contribution by $\gamma - \Pom/\Reg$.
The analogous short-hand notation is used also for other
contributions,  $\Reg_{-} - \Pom$ and $\Ode - \Pom$.

In Fig.~\ref{fig:aux} we compare different processes
for the $pp \to pp \gamma$ reaction
calculated for $\sqrt{s} = 13$~TeV, 
and a somewhat larger range of photon rapidities $|{\rm y}| < 4$, 
and for two $k_{\perp}$ intervals as
specified in the figure legends.
We see that the photoproduction 
is the dominant process for $k_{\perp} > 10$~MeV,
its cross section $d\sigma/dk_{\perp}$
gradually increases and reaches a maximum 
at $k_{\perp} \simeq 0.25$~GeV.
The diffractive bremsstrahlung contribution
is most important in the area of small $k_{\perp}$
but its cross section decreases with increasing $k_{\perp}$.
The other purely diffractive contributions,
$\Reg_{-} - \Pom$ and $\Ode - \Pom$,
give much smaller cross sections.

As was mentioned in \cite{Lebiedowicz:2022nnn} 
the intermediate protons 
in the bremsstrahlung-type diagrams are off shell
when the final state photon is emitted 
from an external proton line. 
But in our model of the bremsstrahlung contributions
we set possible form factors for off-shell protons
in the vertices and in the proton propagator to 1.
We expect that up to $k_{\perp} \simeq 0.1$~GeV
and small $\omega$
the off-shell effects should be small.\footnote{We refer 
the reader to Figs.~8 and 17 of \cite{Lebiedowicz:2022nnn}
where the results for $k_{\perp}$ up to 0.4~GeV are shown.}
Taking this into account,
we do not show here results for the bremsstrahlung mechanism
for a larger $k_{\perp}$ range
where our estimates are uncertain.

It is a known fact that absorption effects due to
strong proton-proton interactions have much more influence
on the purely diffractive processes than on the photoproduction processes; 
see e.g. Table~II of \cite{Lebiedowicz:2019boz}
where the ratios of full and Born cross sections 
for the $pp \to pp \phi$ reaction
for the $\gamma - \Pom$- and $\Ode - \Pom$-fusion processes
are shown.
Thus, it can be expected that 
for the $pp \to pp \gamma$ reaction
at not too large $k_{\perp}$ 
the absorption is small for the $\gamma - \Pom/\Reg$-fusion processes.

It is interesting to note that 
the $\gamma - \Pom/\Reg$-fusion processes 
are of the same order in $\alpha_{\rm em}$
as the bremsstrahlung-type processes via the photon exchange,
which we call the QED bremsstrahlung; see Table~\ref{Table1}.
In the case of the photon production via QED bremsstrahlung
$d\sigma/dt_{1,2}$ increases for $t_{1,2} \to 0$ 
due to the photon propagator which is proportional to $1/t$.
We have checked that for the kinematics considered in our paper
the QED-bremsstrahlung cross section is about
a factor of 200 smaller 
than the diffractive bremsstrahlung cross section
via the pomeron exchange.
For these two bremsstrahlung mechanisms, 
the shapes of the distributions in
${\rm y}$, $k_{\perp}$, and $\omega$ are similar.

\begin{figure}[!ht]
\includegraphics[width=0.49\textwidth]{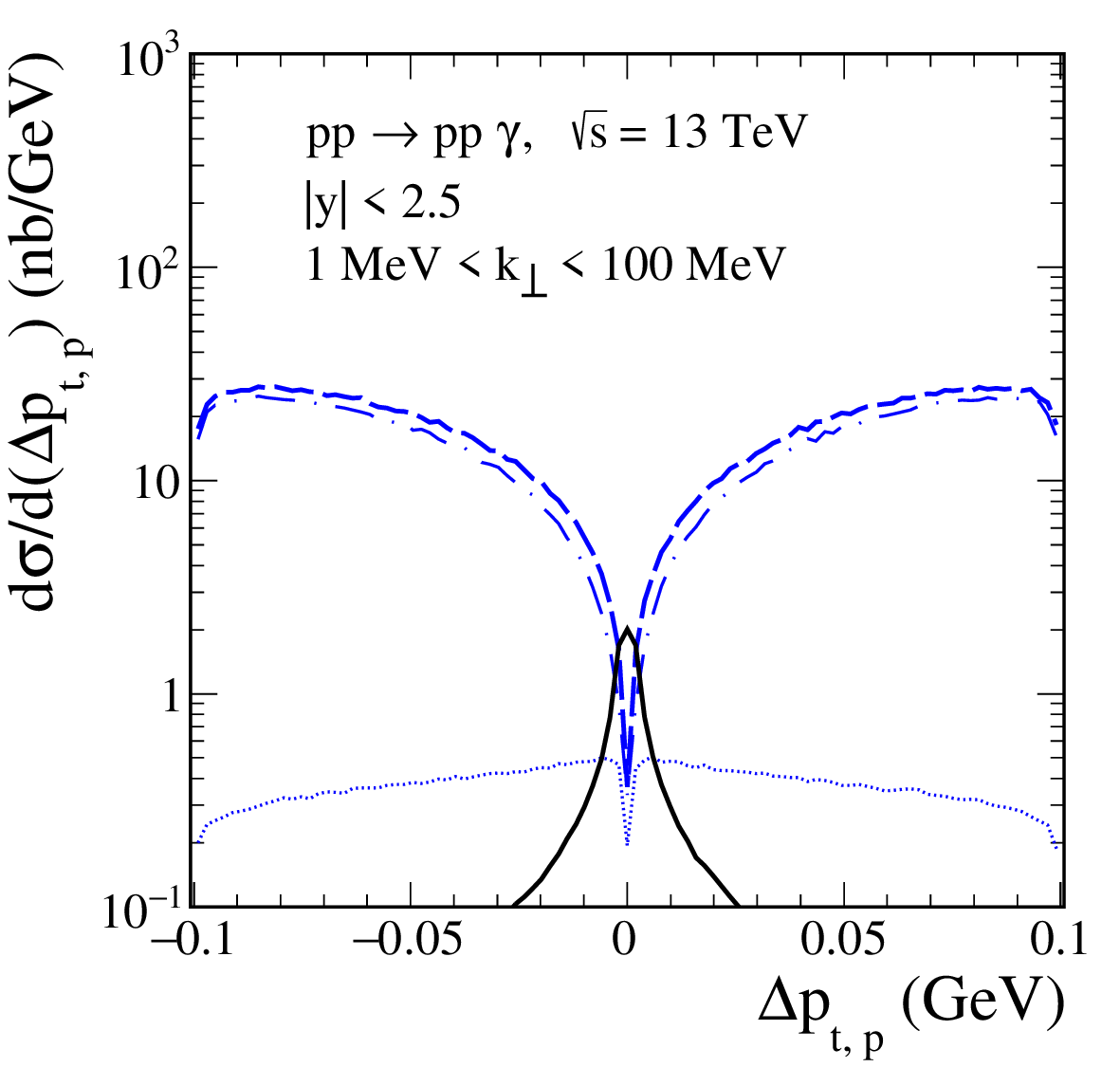}
\includegraphics[width=0.49\textwidth]{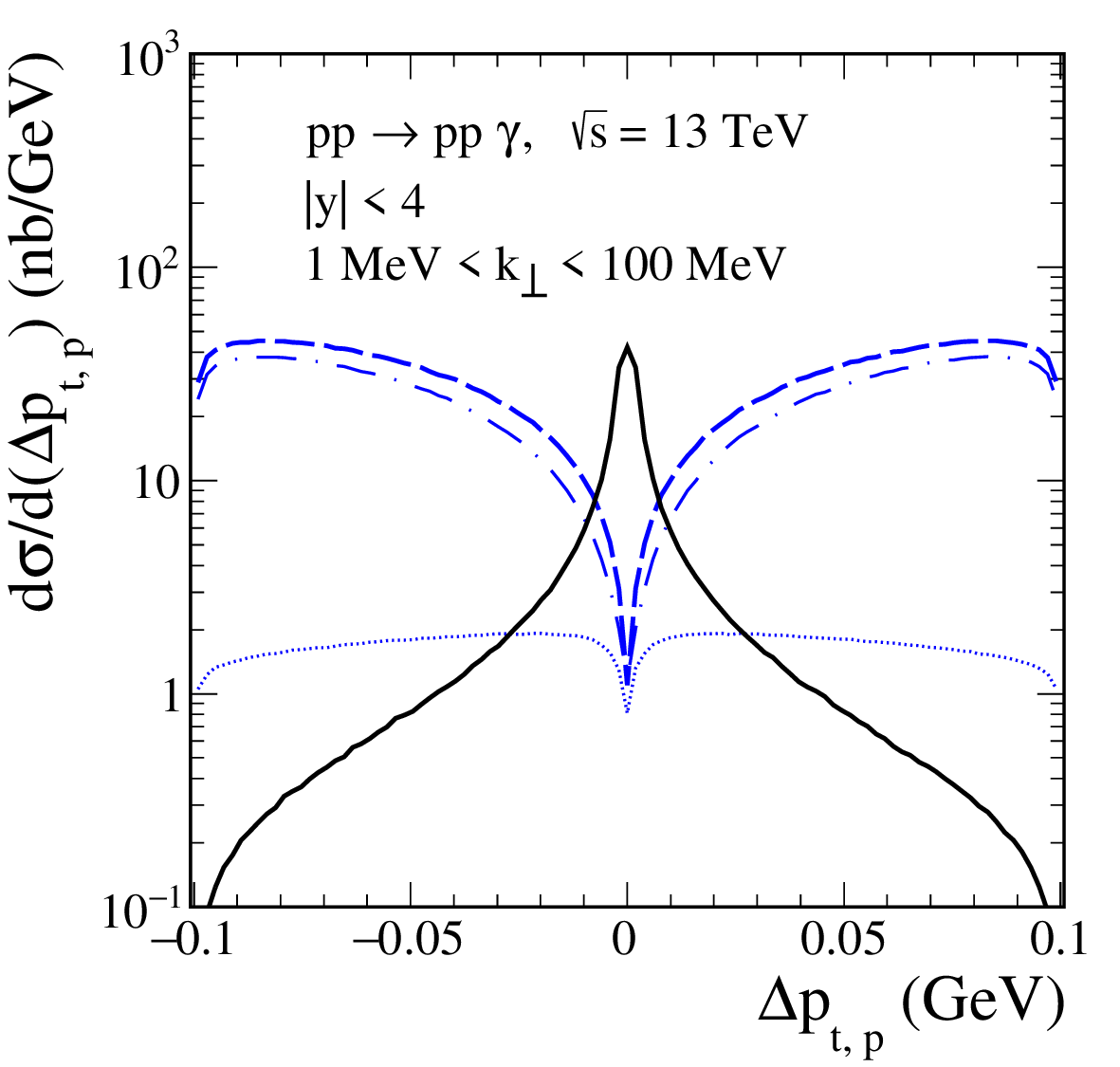}
\caption{\label{fig:pt12}
\small
The distributions in $\Delta p_{t,p}$ 
for the $\gamma - \Pom/\Reg$-fusion processes
(blue lines) 
and diffractive bremsstrahlung (black lines).
The meaning of the lines is the same as in
Fig.~\ref{fig:aux}.
The calculations were done for $\sqrt{s} = 13$~TeV,
$1~{\rm MeV} < k_{\perp} < 100$~MeV,
and for two $|{\rm y}|$ intervals as
specified in the figure legends.}
\end{figure}
In Fig.~\ref{fig:pt12} we present the distributions
in $\Delta p_{t,p} = |\bptap| - |\bptbp|$
for the low-$k_{\perp}$ region 
and for two $|{\rm y}|$ intervals
for the $pp \to pp \gamma$ reaction.
We compare the $\gamma - \Pom/\Reg$-fusion processes
to diffractive bremsstrahlung-type emission of photons
discussed in \cite{Lebiedowicz:2022nnn}.
We see from Figs.~\ref{fig:aux} and \ref{fig:pt12}
that these contributions have 
different characteristics in 
$k_{\perp}$ and $\Delta p_{t,p}$
and that the relative size of the cross sections
depends on the photon-rapidity range.
The~physics behind the results shown in Fig.~\ref{fig:pt12}
is as follows. In the discussion of the results of 
Fig.~\ref{fig:6}
we saw that in the CEP process we have either
$|\bptap| \approx 0$, $|\bptbp|$ sizeable, or
$|\bptbp| \approx 0$, $|\bptap|$ sizeable.
This explains the double-hump structure of the CEP curves
in Fig.~\ref{fig:pt12}.
For bremsstrahlung, on the other hand,
the kinematics of the $pp \to pp \gamma$ reaction
is close to that for elastic scattering, $pp \to pp$,
where $|\bptap| = |\bptbp|$.
We expect, therefore, also for bremsstrahlung 
$|\bptap| \approx |\bptbp|$ and, thus, $\Delta p_{t,p} \approx 0$.
And this is indeed what we see from Fig.~\ref{fig:pt12}.
Furthermore, we see from Fig.~\ref{fig:aux},
upper left panel, that bremsstrahlung increases
relative to CEP for larger $|{\rm y}|$.
And this can also be seen by comparing the left and right panels
of Fig.~\ref{fig:pt12}. 
We conclude by emphasizing that the
measurement of outgoing protons,
necessary to obtain distributions as shown in Fig.~\ref{fig:pt12},
would allow us to better understand
the role of CEP versus bremsstrahlung processes.
Hopefully, the ATLAS and CMS Collaborations will be able to measure photons 
in coincidence with protons in the forward detectors.

\section{Conclusions}
\label{sec:4}

In this paper we have studied 
diffractive production of photons via different fusion processes
within the tensor-pomeron approach
in the reaction $pp \to pp \gamma$ (\ref{pp_ppgam})
at the c.m. energy $\sqrt{s} = 13$~TeV.
We have discussed central-exclusive production (CEP) of photons
from the photoproduction processes
given by the diagrams of Fig.~\ref{fig:pp_pp_gam_CEP_photoprod}
and the purely diffractive fusion processes
given by the diagrams of Fig.~\ref{fig:pp_pp_gam_CEP}.
We can only speak of this CEP mechanism
if the $\gamma p$ subenergies $W_{1}$ and $W_{2}$
[see (\ref{2.17})] are large enough.
We assured this by a cut in the photon rapidity, $|{\rm y}| < 2.5$,
and transverse momentum, $0.1~{\rm GeV} < k_{\perp} < 1$~GeV.

The photoproduction is the dominant CEP mechanism
without considering kinematic cuts on the leading protons.
Due to the virtual photon exchange one of
the protons is scattered at very small angles
(very forward/backward proton rapidities).
The $\gamma - \Pom$ exchanges (\ref{gamP})
populate preferentially the midrapidity region.
The subleading $\gamma - \Reg_{+}$ exchanges
play a role at more forward/backward photon rapidities,
when the energies of the subprocesses
$\gamma^{*} p \to \gamma p$ are smaller
than for ${\rm y} \approx 0$; see Fig.~\ref{fig:2}.
Both, $\Pom$ and $\Reg_{+} = f_{2 \Reg}, a_{2 \Reg}$ exchange
are treated as effective tensor exchanges in our model.
The complete photoproduction result indicates a constructive
interference effect of $\gamma - \Pom$- 
and $\gamma - \Reg_{+}$-fusion processes.
There is, on the other hand, sizeable destructive interference 
between the two terms $\gamma \Pom/\Reg$ and $\Pom/\Reg \gamma$
corresponding to 
the diagrams shown in Fig.~\ref{fig:pp_pp_gam_CEP_photoprod}(a)
and Fig.~\ref{fig:pp_pp_gam_CEP_photoprod}(b), respectively;
see Fig.~\ref{fig:3}.

We have compared the CEP processes to standard results for
bremsstrahlung-type emission of soft photons
discussed previously in \cite{Lebiedowicz:2022nnn}.
We have shown that the photoproduction contribution 
wins over the bremsstrahlung one
for $|\rm y| < 4$ and $k_{\perp} \gtrsim 10$~MeV
(see Fig.~\ref{fig:aux}).
The cross section $d\sigma/dk_{\perp}$
for the $\gamma - \Pom/\Reg$-fusion processes
gradually increases and reaches a maximum 
at $k_{\perp} \simeq 0.25$~GeV.
The purely diffractive CEP fusion processes,
$\Reg_{-} - \Pom$ and $\Ode - \Pom$,
give much smaller cross sections there.
For the photon bremsstrahlung, 
discussed in \cite{Lebiedowicz:2022nnn}, 
the cross section for $k_{\perp} \to 0$ 
diverges as $1/k_{\perp}$.

To summarize: in this article we have studied
central-exclusive production of single photons
in proton-proton collisions at LHC energies, $\sqrt{s} = 13$~TeV.
The CEP process with $\gamma - \Pom$ fusion
is the most important one and preferentially produces photons
at midrapidity.
In this kinematic region CEP dominates over photons
produced by diffractive bremsstrahlung.
And this is true despite the fact that CEP
with $\gamma - \Pom$ fusion is of higher order in $\alpha_{\rm em}$
compared to diffractive bremsstrahlung.
It is clear that our studies can easily be adapted
to $pp$ collisions at RHIC energies, $\sqrt{s} = 200-510$~GeV.

We hope that our theoretical studies
of the $pp \to pp \gamma$ reaction 
will find experimental counterparts by measurements
of soft photons at RHIC and at the LHC.
We emphasize that for detailed comparisons of our predictions
with experiment measurement of the outgoing protons would be
most welcome.

\acknowledgments
This study was partially supported by the Polish National Science Centre
under Grant No. 2018/31/B/ST2/03537.
P.L. was supported by the Bekker Program of the Polish National Agency 
for Academic Exchange,
Project No. BPN/BEK/2021/2/00009/U/00001.

\appendix
\section{Effective propagator and vertex functions}
\label{sec:appendixA}

We give here a list of all effective propagator 
and vertex functions used in the calculation of
the $\gamma - \Pom/\Reg$-fusion processes discussed 
in Sec.~\ref{sec:2B}.

For the two tensor pomerons, soft $(j=1)$ and hard $(j=0)$,
we use the effective propagators and the $\Pom_{j} pp$ vertices
as given in Appendix~A of \cite{Britzger:2019lvc},
\begin{eqnarray}
&&i\Delta^{(\Pom_{j})}_{\mu \nu, \kappa \lambda}(s,t) = 
\frac{1}{4s} \left( g_{\mu \kappa} g_{\nu \lambda} 
                  + g_{\mu \lambda} g_{\nu \kappa}
                  - \frac{1}{2} g_{\mu \nu} g_{\kappa \lambda} \right)
(-i s \tilde{\alpha}'_{j})^{\alpha_{j}(t)-1}\,,
\label{A1}\\
&&i\Gamma_{\mu \nu}^{(\Pom_{j} pp)}(p',p)
=-i 3 \beta_{jpp} F_{1}^{(j)}(t)
\left\lbrace 
\frac{1}{2} 
\left[ \gamma_{\mu}(p'+p)_{\nu} 
     + \gamma_{\nu}(p'+p)_{\mu} \right]
- \frac{1}{4} g_{\mu \nu} (\slash{p}' + \slash{p})
\right\rbrace.\qquad
\label{A2}
\end{eqnarray}
Here $\beta_{1 pp} = \beta_{0 pp} = 1.87$~GeV$^{-1}$
are coupling constants and
$F_{1}^{(j)}(t)$ are form factors 
[see Eq.~(\ref{A5}) below].
The \textit{Ans{\"a}tze} 
for effective propagators and vertices
for the tensor reggeons
$f_{2 \Reg}$ and $a_{2 \Reg}$ have the same structure
as (\ref{A1}) and (\ref{A2}), respectively.
The contributions of these reggeons 
are combined into one term $\Reg_{+}$, labeled $j = 2$, 
and the coupling constant for $\Reg_{+} pp$ is given as
$\beta_{2 pp} = 3.68$~GeV$^{-1}$;
see (A29) of \cite{Britzger:2019lvc}.
The pomeron and reggeon trajectory functions
are assumed to be of linear form
\begin{eqnarray}
\alpha_{j}(t) = \alpha_{j}(0)+\alpha'_{j}\,t\,,
\quad
\alpha_{j}(0) = 1 + \epsilon_{j}\,,
\quad
j = 0, 1, 2\,.
\label{trajectories}
\end{eqnarray}
The values of the intercept parameters of the Regge trajectories 
obtained in \cite{Britzger:2019lvc} from a comparison to HERA DIS data
are
\begin{eqnarray}
\epsilon_{1} = 0.0935 (^{+76}_{-64})\,, \quad 
\epsilon_{0} = 0.3008  (^{+73}_{-84})\,, \quad
\alpha_{2}(0) = 0.485 (^{+88}_{-90})\,.
\label{intercept_parameters}
\end{eqnarray}
For the slope parameters default values
were used in \cite{Britzger:2019lvc}
$\alpha'_{1} = \alpha'_{0} = 0.25\;{\rm GeV}^{-2}$, and
$\alpha'_{2} = 0.9\;{\rm GeV}^{-2}$,
and the scale parameters $\tilde{\alpha}'_{j}$
were chosen equal to the slope parameters $\alpha_{j}'$.
In our present work, in the calculation of
the $\gamma - \Pom/\Reg$ CEP process,
we use the above central values
for $\epsilon_{1}$, $\epsilon_{0}$, and $\alpha_{2}(0)$
and the default values for $\alpha_{j}'$ ($j = 0, 1, 2$).
In the calculation of the purely diffractive contributions,
discussed in Sec.~\ref{sec:2C},
for $\epsilon_{1}$ we use the value $\epsilon_{1} = 0.0865$
as determined by us in \cite{Lebiedowicz:2022nnn}.
This latter value is about 1 s.d. lower than the value from
(\ref{intercept_parameters}).
The default value from \cite{Donnachie:2002en, Ewerz:2013kda} 
is $\epsilon_{1} = 0.0808$
which is 2 s.d. lower than the one from (\ref{intercept_parameters}).
We can motivate our use of different values of $\epsilon_{1}$
for the different processes shown in 
Fig.~\ref{fig:pp_pp_gam_CEP_photoprod}
and Fig~\ref{fig:pp_pp_gam_CEP} as follows.
In the diagrams of Fig.~\ref{fig:pp_pp_gam_CEP_photoprod}
the pomeron exchange is in the subreaction $\gamma^{*}p \to \gamma p$
and we think that it is appropriate to use there the value of $\epsilon_{1}$
from (\ref{intercept_parameters}) as determined from the closely
related process $\gamma^{*}p \to \gamma^{*}p$; see \cite{Britzger:2019lvc}.
In Fig~\ref{fig:pp_pp_gam_CEP}, on the other hand, we have a diffractive
collision being close to $pp$ elastic scattering where
we determined a slightly lower value of $\epsilon_{1}$ 
in \cite{Lebiedowicz:2022nnn}.
We could, for instance, consider this slightly lower value of $\epsilon_{1}$
in the hadronic diffraction case as being due to stronger absorption effects
there compared to the $\gamma^{*}p \to \gamma^{*}p$ reaction.
In any case, in the kinematical range considered by us here
the diffractive contribution to the CEP of a single photon is small;
see Fig.~\ref{fig:aux}.

The \textit{Ans{\"a}tze} for the $\Pom_{j} \gamma^{*} \gamma^{*}$ and
$\Reg_{+}\gamma^{*} \gamma^{*}$ 
coupling functions for both real and virtual photons
are given in \cite{Britzger:2019lvc}.
The $\Pom_{j} \gamma^{*} \gamma^{*}$ ($j = 0,1$) vertex reads 
\begin{eqnarray}
i\Gamma^{(\Pom_{j} \gamma^{*} \gamma^{*})}_{\mu \nu \kappa \rho}(q',q) =
i\left[
2a_{j \gamma^{*} \gamma^{*}}(q^{2},q'^{2},t)\,\Gamma^{(0)}_{\mu \nu \kappa \rho}(q',-q)\,
     - b_{j \gamma^{*} \gamma^{*}}(q^{2},q'^{2},t)\,\Gamma^{(2)}_{\mu \nu \kappa \rho}(q',-q)
     \right], \qquad
\label{A3}
\end{eqnarray}  
where $t = (q - q')^{2}$.
The rank-4 tensor functions 
are defined in 
(A13) and (A14) of \cite{Britzger:2019lvc},
\begin{eqnarray}
\label{A3a}
\Gamma_{\mu\nu\kappa\lambda}^{(0)} (k_1,k_2) &=& 
[(k_1\cdot k_2) g_{\mu\nu} - k_{2\mu} k_{1\nu}] 
\left[k_{1\kappa}k_{2\lambda} + k_{2\kappa}k_{1\lambda}
- \frac{1}{2} (k_1 \cdot k_2) g_{\kappa\lambda}\right] \,,
\\
\label{A3b}
\Gamma_{\mu\nu\kappa\lambda}^{(2)} (k_1,k_2) &=&
(k_1\cdot k_2) (g_{\mu\kappa} g_{\nu\lambda} + g_{\mu\lambda} g_{\nu\kappa} )
+ g_{\mu\nu} (k_{1\kappa} k_{2\lambda} + k_{2\kappa} k_{1\lambda} ) 
\nonumber\\
&& - k_{1\nu} k_{2 \lambda} g_{\mu\kappa} - k_{1\nu} k_{2 \kappa} g_{\mu\lambda} 
- k_{2\mu} k_{1 \lambda} g_{\nu\kappa} - k_{2\mu} k_{1 \kappa} g_{\nu\lambda} 
\nonumber\\
&& - [(k_1 \cdot k_2) g_{\mu\nu} - k_{2\mu} k_{1\nu} ] \,g_{\kappa\lambda} \,.
\end{eqnarray}  
The $\Reg_{+} \gamma^{*} \gamma^{*}$ vertex
for real and virtual photons has the same structure 
as shown in (\ref{A3}) with $j = 2$;
see (A27)--(A31) of \cite{Britzger:2019lvc}.
The coupling functions $a_{j \gamma^{*} \gamma^{*}}(q^{2},q'^{2},t)$ 
and $b_{j \gamma^{*} \gamma^{*}}(q^{2},q'^{2},t)$,
for the case $q^{2} = -Q^{2}$, $q'^{2} = 0$,
are taken as in
(2.21)--(2.23) of \cite{Lebiedowicz:2022xgi},
\begin{eqnarray}
&&a_{j \gamma^{*} \gamma^{*}}(q^{2},0,t) = 
e^{2} \hat{a}_{j}(Q^{2}) F^{(j)}(t)\,, \quad j = 0, 1, 2\,,
\nonumber \\
&&b_{2 \gamma^{*} \gamma^{*}}(q^{2},0,t) = 
e^{2} \hat{b}_{2}(Q^{2}) F^{(2)}(t)\,.
\label{A4}
\end{eqnarray}  
For two alternative fits for $b_{1 \gamma^{*} \gamma^{*}}$
and $b_{0 \gamma^{*} \gamma^{*}}$
we obtained from a comparison to HERA DVCS data
\begin{eqnarray}
{\rm FIT~1}: \;
&&b_{1 \gamma^{*} \gamma^{*}}(q^{2},0,t) = e^{2} \hat{b}_{1}(0) 
(1 + Q^{2}/\Lambda_{1}^{2})^{-1.2}\, F^{(1)}(t)\,,
\quad \Lambda_{1} = 1.4\;{\rm GeV}\,, \nonumber\\
&&b_{0 \gamma^{*} \gamma^{*}}(q^{2},0,t) = e^{2} \hat{b}_{0}(Q^{2}) 
F^{(0)}(t)\,,
\label{FIT1} \\
{\rm FIT~2}:\;
&&b_{1 \gamma^{*} \gamma^{*}}(q^{2},0,t) = e^{2} \hat{b}_{1}(0) 
(1 + Q^{2}/\Lambda_{2}^{2})^{-2.0} \, F^{(1)}(t)\,,
\quad \Lambda_{2} = 2.0\;{\rm GeV}\,, \nonumber \\
&&
b_{0 \gamma^{*} \gamma^{*}}(q^{2},0,t) =
\left\{ \begin{array}{lr} 
e^{2} \hat{b}_{0}(Q^{2}) \, F^{(0)}(t)
& {\rm for}\ Q^{2} < 1.5~{\rm GeV}^{2} \\ 
e^{2} \Lambda_{0} (1 + Q^{2}/\Lambda_{3}^{2})^{-0.6} \, F^{(0)}(t)
& {\rm for}\ Q^{2} \geq 1.5~{\rm GeV}^{2} 
\end{array}\right. \,, \nonumber \\
&&\Lambda_{0} = 9.46 \times 10^{-3} \; {\rm GeV}^{-1}\,, \quad 
\Lambda_{3} = 2.3\;{\rm GeV}.
\label{FIT2}
\end{eqnarray}
See (2.21)--(2.23) of \cite{Lebiedowicz:2022xgi}.
The coupling functions 
$\hat{a}_{j}(Q^{2})$ and $\hat{b}_{j}(Q^{2})$ 
were determined in \cite{Britzger:2019lvc}
from the global fit to HERA inclusive DIS data
for $Q^{2} < 50$~GeV$^{2}$ and $x < 0.01$
and the ($Q^{2} = 0$) photoproduction data.
According to \cite{Britzger:2019lvc}, 
for the $R_{+}$-reggeon term, 
$\hat{a}_{2}(Q^{2}) = 0$ while the function $\hat{b}_{2}(Q^{2})$
vanishes rapidly with increasing $Q^{2}$.
All coupling functions $\hat{a}_{j}$ and $\hat{b}_{j}$ 
are plotted in Fig.~2 of \cite{Lebiedowicz:2022xgi}.
For small $Q^{2}$, the soft pomeron function $b_{1 \gamma^{*} \gamma^{*}}$
gives a larger contribution to the cross section
than the corresponding hard one $b_{0 \gamma^{*} \gamma^{*}}$.
In the large $Q^{2}$ region the reverse is found.

We use the combined form-factor functions
for a given $j$~$(j = 0, 1, 2)$
\begin{equation}
F_{\rm eff}^{(j)}(t) = 
F^{(j)}(t) \times  F_{1}^{(j)}(t) =
\exp(-b_{j}|t|/2 )\,,
\label{A5}
\end{equation}
assuming the same $t$ dependence for both $a$ and $b$ 
coupling functions.
We take
$b_{1} = b_{2} = 5.0\; {\rm GeV}^{-2}$
and 
$b_{0} = 1.0\; {\rm GeV}^{-2}$ from \cite{Lebiedowicz:2022xgi}.

\section{Estimate of the cross section for $pp \to pp \gamma$ using the method of the equivalent photon flux}
\label{sec:appendixB}

In this appendix we estimate the cross section for
the $\gamma \Pom$ fusion contribution to $pp \to pp \gamma$
using the equivalent photon-flux method.
We consider the diagram of Fig.~\ref{fig:pp_pp_gam_CEP_photoprod}~(a).
The main contribution comes from the region 
where the absolute value of the invariant mass squared $t_{1}$
of the exchanged photon $\gamma^{*}$ is very small;
see Fig.~\ref{fig:6}(a).
In the following we work in the overall c.m. system of the reaction
(\ref{pp_ppgam}) where we have
\begin{eqnarray}
&& p_{a}^{0} = p_{b}^{0} = \frac{1}{2} \sqrt{s}\,, \nonumber \\
&& p_{1}'^{0} = \frac{1}{2\sqrt{s}} (s+m_{p}^{2}-s_{2})\,.
\label{B1}
\end{eqnarray}
With $\vartheta$ the angle between $\bpaap$ and $\bpa$ we get
\begin{eqnarray}
|t_{1}| &=& 2 (p_{a}, p_{1}') - 2 m_{p}^{2}\nonumber \\
        &=& 2 p_{a}^{0} p_{1}'^{0} - 2 m_{p}^{2} - 2 |\bpa| |\bpaap| \cos\vartheta \,,
\label{B2}
\end{eqnarray}
having the minimal value
\begin{eqnarray}
|t_{1}|_{\rm min} &=& 2 (p_{a}^{0} p_{1}'^{0} - m_{p}^{2} - |\bpa| |\bpaap|) \nonumber \\
                  &=& \frac{(s_{2}-m_{p}^{2})^{2} m_{p}^{2}}{2 s (p_{a}^{0} p_{1}'^{0} - m_{p}^{2} + |\bpa| |\bpaap|)}\,.
\label{B3}
\end{eqnarray}
For our case we have (see Fig.~\ref{fig:2})
\begin{eqnarray}
&&\sqrt{s} = 13 \; {\rm TeV}\,,\quad
\sqrt{s} \gg W_{2} = \sqrt{s_{2}} \geqslant 10 \; {\rm GeV}\,, \nonumber \\
&&|t_{1}|_{\rm min} \cong \left( \frac{W_{2}}{\sqrt{s}}\right)^{4} m_{p}^{2}\,.
\label{B4}
\end{eqnarray}
Numerically we get
\begin{eqnarray}
&&|t_{1}|_{\rm min} \cong 3.1 \times 10^{-13}\; {\rm GeV}^{2}
\quad {\rm for} \; W_{2} = 10 \; {\rm GeV}\,, \nonumber \\
&&|t_{1}|_{\rm min} \cong 4.9 \times 10^{-8}\; {\rm GeV}^{2}
\quad {\rm for} \; W_{2} = 200 \; {\rm GeV}\,.
\label{B5}
\end{eqnarray}
These are very small values for $|t_{1}|$ where
the equivalent photon method should give good estimates.

The equivalent photon fluxes of various particles
are listed in Appendix~D of \cite{Budnev:1975poe}.
For the proton this flux reads for our case
\begin{eqnarray}
dn(q_{1}^{0},t_{1}) = \frac{\alpha_{\rm em}}{\pi} 
\frac{dq_{1}^{0}}{q_{1}^{0}} \frac{d|t_{1}|}{|t_{1}|} 
\left[ \left(1 - \frac{q_{1}^{0}}{p_{a}^{0}} \right) D(t_{1})
+ \frac{1}{2} \left( \frac{q_{1}^{0}}{p_{a}^{0}} \right)^{2} C(t_{1})
- \left(1 - \frac{q_{1}^{0}}{p_{a}^{0}} \right) \frac{|t_{1}|_{\rm min}}{|t_{1}|}
D(t_{1}) \right]. \nonumber\\
\label{B6}
\end{eqnarray}
Here
\begin{eqnarray}
C(t) &=& G_{M}^{2}(t)\,,\nonumber \\
D(t) &=& [4 m_{p}^{2} G_{E}^{2}(t) - t \,G_{M}^{2}(t)]
[4 m_{p}^{2} - 1]^{-1}\,,
\label{B7}
\end{eqnarray}
and $G_{E}$ and $G_{M}$ are the electric and magnetic form factor
of the proton, respectively,
\begin{eqnarray}
G_{E}(t) &=& F_{1}(t) + \frac{t}{4 m_{p}^{2}} F_{2}(t) \,,\nonumber \\
G_{M}(t) &=& F_{1}(t) + F_{2}(t) \,.
\label{B8}
\end{eqnarray}
In our case we have
\begin{eqnarray}
&&W_{2}^{2} \cong 2 q_{1}^{0} \sqrt{s}\,, \nonumber\\
&&\frac{q_{1}^{0}}{p_{a}^{0}} \cong \frac{W_{2}^{2}}{s} \ll 1\,, \nonumber\\
&&\frac{dq_{1}^{0}}{q_{1}^{0}} = \frac{2 dW_{2}}{W_{2}}\,.
\label{B9}
\end{eqnarray}
We can, thus, neglect in (\ref{B6}) the terms with
$q_{1}^{0}/p_{a}^{0}$ and $(q_{1}^{0}/p_{a}^{0})^{2}$
and set $D(t_{1}) \approx 1$.
We~get then as an estimate for the cross section via the $\gamma \Pom$ fusion
\begin{eqnarray}
\left.d\sigma(pp \to pp \gamma)\right|_{\gamma \Pom \, {\rm fusion}} &\cong&
dn(q_{1}^{0},t_{1}) \sigma_{\rm T}^{\gamma p \to \gamma p}(t_{1} = 0, W_{2} = \sqrt{2 q_{1}^{0} \sqrt{s}}) \nonumber \\
&=&
\frac{2 \alpha_{\rm em}}{\pi} 
\frac{dW_{2}}{W_{2}} \frac{d|t_{1}|}{|t_{1}|}
\left(1 - \frac{|t_{1}|_{\rm min}}{|t_{1}|} \right)
\sigma_{\rm T}^{\gamma p \to \gamma p}(0, W_{2})\,.
\label{B10}
\end{eqnarray}
Here $\sigma_{\rm T}^{\gamma p \to \gamma p}(0, W)$
is the total cross section for Compton scattering
of a real photon on the proton at c.m. energy $W$.
Now we integrate (\ref{B10}) over $|t|$ from $|t_{1}|_{\rm min}$
to a maximal value $|t_{1}|_{\rm max}$ which we set
to 1~GeV$^{2}$.
We get then
\begin{eqnarray}
\left.\frac{d\sigma(pp \to pp \gamma)}{dW_{2}}\right|_{\gamma \Pom \, {\rm fusion}} 
&\cong&
\frac{2 \alpha_{\rm em}}{\pi} 
\frac{1}{W_{2}} \int_{|t_{1}|_{\rm min}}^{|t_{1}|_{\rm max}} 
\frac{d|t_{1}|}{|t_{1}|}
\left(1 - \frac{|t_{1}|_{\rm min}}{|t_{1}|} \right)
\sigma_{\rm T}^{\gamma p \to \gamma p}(0, W_{2}) \nonumber \\
&\cong&
\frac{2 \alpha_{\rm em}}{\pi} 
\frac{1}{W_{2}} 
\left[ \ln \frac{|t_{1}|_{\rm max}}{|t_{1}|_{\rm min}} 
-1 + \frac{|t_{1}|_{\rm min}}{|t_{1}|_{\rm max}} \right]
\sigma_{\rm T}^{\gamma p \to \gamma p}(0, W_{2})\,.
\label{B11}
\end{eqnarray}
With $|t_{1}|_{\rm max} = 1$~GeV$^{2}$ and $|t_{1}|_{\rm min}$
from (\ref{B5}) we get
\begin{eqnarray}
\frac{2 \alpha_{\rm em}}{\pi} 
\left[ \ln \frac{|t_{1}|_{\rm max}}{|t_{1}|_{\rm min}} 
-1 + \frac{|t_{1}|_{\rm min}}{|t_{1}|_{\rm max}} \right]
\cong 
\begin{cases}
0.13 & \mbox{for}\; W_{2} = 10~\mathrm{GeV} \,,\\
0.07 & \mbox{for}\; W_{2} = 200~\mathrm{GeV} \,.\\
\end{cases}
\label{B12}
\end{eqnarray}
That is, this quantity has a slow decrease with $W_{2}$.

From Fig.~3 of \cite{Lebiedowicz:2022xgi} we see that the cross section
$\sigma_{\rm T}^{\gamma p \to \gamma p}(0, W)$
is around 100~nb for $W \simeq 10$~GeV and slowly rising with $W$.
Therefore, we get as an estimate
\begin{eqnarray}
\left.\frac{d\sigma(pp \to pp \gamma)}{dW_{2}}\right|_{\gamma \Pom \, {\rm fusion}} 
\cong
1.3 \,\frac{10}{W_{2}} \;{\rm nb/GeV}
\,.
\label{B13}
\end{eqnarray}
Here $W_{2}$ has to be included in GeV.
Our estimate gives a cross section falling roughly as $1/W_{2}$
with a size of order 0.3~nb/GeV for $W_{2} = 40$~GeV.

We can compare this with the result of the explicit calculation
shown in the right panel of Fig.~\ref{fig:1}.
Since there $W_{1}$ is plotted we must compare with the dashed red line
corresponding to the $\Pom/\Reg \gamma$ fusion.
And, indeed, this cross section is not far from
0.3~nb/GeV for $W_{1} = 40$~GeV
and it is falling with increasing $W_{1}$ according to
a $1/W_{1}$ law.

\section{CEP of photons in the limit $k \to 0$}
\label{sec:appendixC}

In this appendix we investigate the CEP mechanism for photons
in the limit that the four momentum $k$ of the photon approaches zero.
We know from Low's theorem \cite{Low:1958sn} that in this limit
the bremsstrahlungs mechanism will dominate.
But we find it interesting to see how CEP goes over to bremsstrahlung
for $k \to 0$.

\begin{figure}[!h]
(a)\includegraphics[width=6cm]{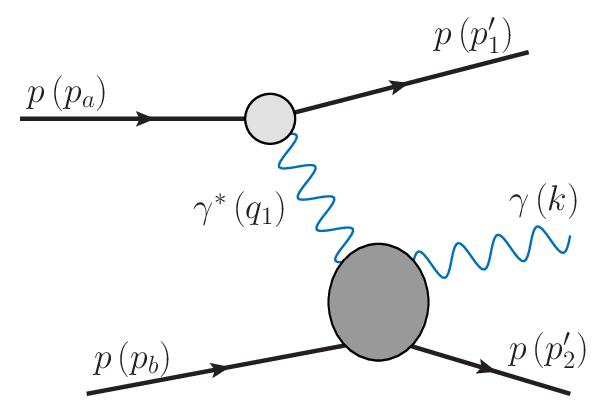}\qquad
(b)\includegraphics[width=5.7cm]{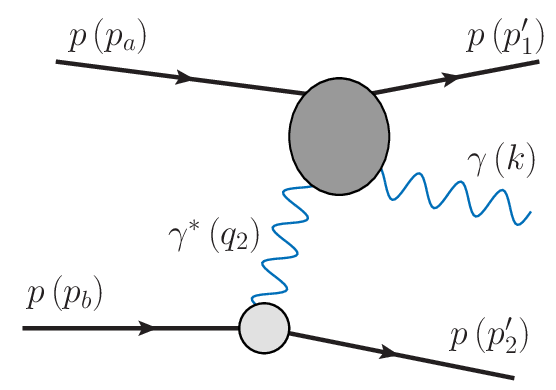}
\caption{
General diagrams for photon production with $\gamma^{*}$ exchange.}
\label{fig:C1}
\end{figure}

Let us start with the diagrams with $\gamma^{*}$ exchange shown in
Fig.~\ref{fig:pp_pp_gam_CEP_photoprod}.
For general $W_{1}$ and $W_{2}$ we write them as emission of $\gamma^{*}$ 
followed by the scattering process $\gamma^{*} p \to \gamma p$
as shown in Fig.~\ref{fig:C1}.

\begin{figure}[!h]
\includegraphics[width=4.7cm]{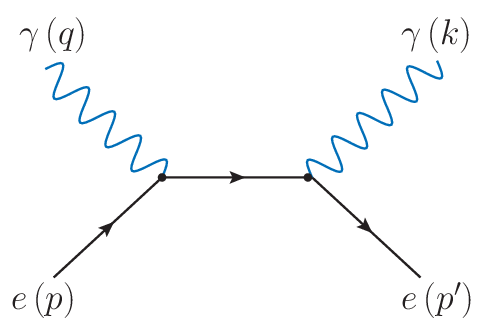}\qquad
\includegraphics[width=4.7cm]{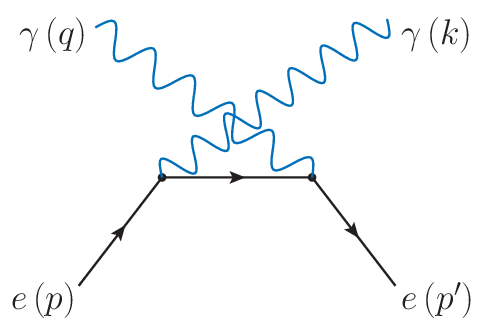}
\caption{The $s$ and $u$ channel diagrams 
for $\gamma e \to \gamma e$.}
\label{fig:C2}
\end{figure}

For $k \to 0$ we have from (\ref{2.17}) 
$W_{1}^{2}, W_{2}^{2} \to m_{p}^{2}$
and the description of the $\gamma^{*} p \to \gamma p$ scattering
by $t$-channel exchanges as shown in Fig.~\ref{fig:pp_pp_gam_CEP_photoprod}
will certainly no longer be adequate.
We note that for real Compton scattering on an electron
\begin{eqnarray}
\gamma (q) + e (p) \to \gamma (k) + e (p')\,,
\label{C2}
\end{eqnarray}
with the initial and final photons on shell, there exists a low energy
theorem in QED proven by W.~Thirring \cite{Thirring:1950cy}.
In the limit $k \to 0$ the amplitude for (\ref{C2}) is given
exactly by the $s$ and $u$ channel diagrams shown in Fig.~\ref{fig:C2}.

\begin{figure}[!h]
\includegraphics[width=6.9cm]{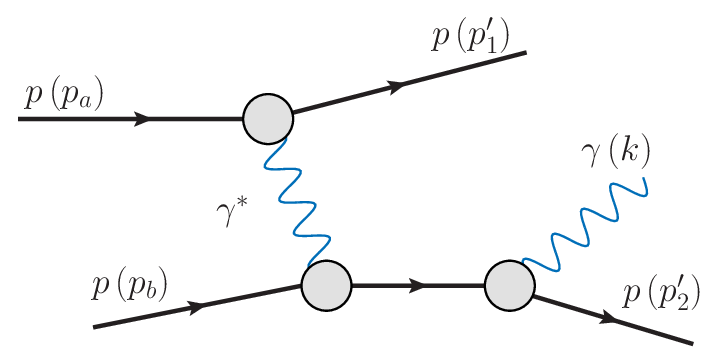}\qquad
\includegraphics[width=5.7cm]{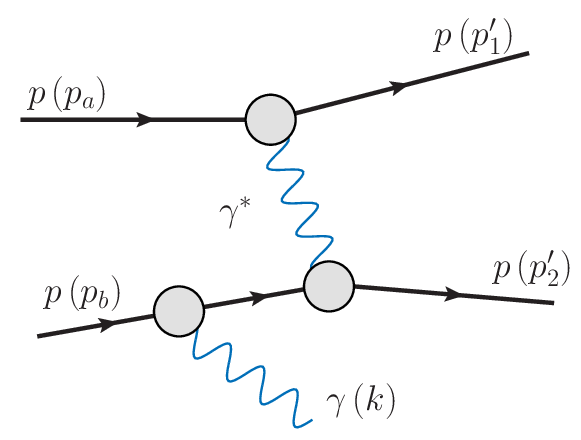}
\caption{Bremsstrahlungs diagrams obtained 
from the diagram Fig.~\ref{fig:C1}(a) for $k \to 0$.
From the diagram Fig.~\ref{fig:C1}(b) we get for $k \to 0$
the analogous diagrams where the final photon $\gamma(k)$
is emitted from the upper proton line.}
\label{fig:C3}
\end{figure}

Thus, we expect that also in our case
where we have in Figs.~\ref{fig:C1}(a) and \ref{fig:C1}(b)
\begin{eqnarray}
\gamma^{*} (q_{1}) + p (p_{b}) \to 
\gamma (k) + p (p_{2}')
\label{C3}
\end{eqnarray}
and
\begin{eqnarray}
\gamma^{*} (q_{2}) + p (p_{a}) \to 
\gamma (k) + p (p_{1}')\,,
\label{C4}
\end{eqnarray}
respectively, the analogous $s$ and $u$ channel diagrams will be important.
Inserting these in the diagrams of Fig.~\ref{fig:C1}(a) and \ref{fig:C1}(b)
we get the diagrams of Fig.~\ref{fig:C3}
which are exactly the bremsstrahlung diagrams associated with
the basic $\gamma^{*}$ exchange.
We know from Low's theorem \cite{Low:1958sn} that for $k \to 0$
these diagrams give the leading behavior proportional to $1/\omega$
in the $pp \to pp \gamma$ amplitude
when the basic diagram for $pp \to pp$ 
is the one with $\gamma^{*}$ exchange.
That is, for $k \to 0$ the diagrams of Fig.~\ref{fig:C1} 
give those of the QED bremsstrahlung process; 
see Table~\ref{Table1}.

To summarize: for high $W_{1}$ and $W_{2}$,
greater than 6~GeV say,
the most important contributions to $\gamma^{*} p \to \gamma p$
in the diagrams of Fig.~\ref{fig:C1} will come
from the $t$-channel exchanges, $\Pom + \Reg_{+}$.
This then leads to the $\gamma - \Pom/\Reg$ CEP process.
For $W_{1,2} \to m_{p}^{2}$, on the other hand,
the diagrams of Fig.~\ref{fig:C1} give the QED bremsstrahlung ones.
For intermediate values of $W_{1,2}$
a simple addition of contributions from
$s$, $u$, and $t$ channel diagrams 
for $\gamma^{*} p \to \gamma p$
is probably not the right thing to do.
We have to recall the duality arguments which were initiated
by the discovery of the Veneziano amplitude \cite{Veneziano:1968yb}.
We think that, therefore, a strict separation of CEP
and bremsstrahlungs-type contributions to $pp \to pp \gamma$
is in general not possible.

The $k \to 0$ behavior of the amplitudes 
of the diffractive contributions to photon CEP, 
see Sec.~\ref{sec:2C} and Fig.~\ref{fig:pp_pp_gam_CEP},
can be discussed in an analogous way.
For $k \to 0$ we shall obtain from 
the analogs of Fig.~\ref{fig:C1}
with $\gamma^{*}$ replaced by hadronic exchanges
the diffractive bremsstrahlungs diagrams for $pp \to pp \gamma$ shown in
Figs.~3(a)--3(f) of \cite{Lebiedowicz:2022nnn}.
There only $\Pom$ exchange is shown but its replacement
by the other hadronic exchanges,
$f_{2 \Reg}$, $a_{2 \Reg}$, $\Ode$, $\omega_{\Reg}$, $\rho_{\Reg}$,
is mentioned in the figure caption.

\bibliography{refs}

\end{document}